\documentclass[a4paper,11pt]{article}
\pdfoutput=1 
\usepackage{jcappub}
\usepackage{amsmath}
\usepackage{graphicx}
\usepackage{latexsym}
\usepackage{xspace}
\usepackage{color}
\usepackage{hyperref} 
\usepackage{bm}
\usepackage{relsize}
\usepackage{tabularx}
\usepackage{multirow}
\usepackage{amssymb}
\usepackage[table]{xcolor}
\usepackage{slashbox}
\usepackage{braket}
\usepackage{comment}
\usepackage[utf8]{inputenc}
\usepackage[normalem]{ulem}

\makeatletter
\gdef\@fpheader{}
\g@addto@macro\bfseries{\boldmath}
\makeatother

\newcommand{\dif}[2]{\frac{\mathrm{d} #1}{\mathrm{d} #2}}
\newcommand{\pdif}[2]{\frac{\partial #1}{\partial #2}}
\newcommand{\ppdif}[3]{\frac{\partial^2 #1}{\partial #2\partial #3}}
\newcommand{\bae}[1]{\begin{align} #1 \end{align}}

\newcommand{\FNC}{\overline{\mathrm{FNC}}}
\newcommand{\kL}{k_\uL }
\newcommand{\kS}{k_\uS }

\newcommand{\ltsim}{\protect\raisebox{-0.5ex}{$\:\stackrel{\textstyle <}{\sim}\:$}}
\newcommand{\gtsim}{\protect\raisebox{-0.5ex}{$\:\stackrel{\textstyle >}{\sim}\:$}}

\newcommand{\ie}{{i.e.~}}

\newcommand{\eg}{\textsl{e.g.~}}

\newcommand{\order}[1]{\mathcal{O}\!\left(#1\right)}

\newcommand{\dd}{\mathrm{d}}
\newcommand{\ee}{e}

\newcommand{\sss}[1]{{\scriptscriptstyle{#1}}}

\newcommand{\uPl}{\mathrm{Pl}}
\newcommand{\uin}{\mathrm{in}}

\newcommand{\uc}{\mathrm{c}}

\newcommand{\uS}{\mathrm{S}}

\newcommand{\uL}{\mathrm{L}}
\newcommand{\uB}{\mathrm{B}}
\newcommand{\uC}{\mathrm{C}}

\newcommand{\usssS}{\sss{\uS}}

\newcommand{\usssPl}{\sss{\uPl}}

\newcommand{\nS}{n_\usssS}

\newcommand{\uNL}{\mathrm{NL}}

\newcommand{\calP}{\mathcal{P}}

\newcommand{\GeV}{\mathrm{GeV}}

\newcommand{\cs}{c_{_\mathrm{S}}}
\newcommand{\dotcs}{\dot{c}_{_\mathrm{S}}}

\newcommand{\Mp}{M_\usssPl}

\newcommand{\fnl}{f_\uNL}

\newcommand{\efolds}{$e$-folds~}

\newcommand{\beq}{\begin{equation}}
\newcommand{\eeq}{\end{equation}}
\newcommand{\bea}{\begin{eqnarray}}
\newcommand{\eea}{\end{eqnarray}}

\newlength{\wsingfig}
\newlength{\wdblefig}
\newlength{\wquadfig}
\newlength{\wtriplefig}
\newcommand{\Eq}[1]{Eq.~(\ref{#1})}
\newcommand{\Eqs}[1]{Eqs.~(\ref{#1})}
\newcommand{\Fig}[1]{Fig.~{\ref{#1}}}
\newcommand{\Figs}[1]{Figs.~{\ref{#1}}}
\newcommand{\Ref}[1]{Ref.~{\cite{#1}}}
\newcommand{\Refs}[1]{Refs.~{\cite{#1}}}
\newcommand{\Sec}[1]{Sec.~\ref{#1}}

\newcommand{\App}[1]{Appendix~\ref{#1}}
\newcommand{\Apps}[1]{Appendices~\ref{#1}}

\subheader{}

\title{Squeezed Bispectrum in the $\delta N$ Formalism: Local Observer Effect in Field Space}

\author[a,b]{Yuichiro Tada}
\author[c]{and Vincent Vennin}

\affiliation[a]{Kavli Institute for the Physics and Mathematics of the Universe (WPI), UTIAS, The University of Tokyo, 
5-1-5 Kashiwanoha, Kashiwa, Chiba 277-8583, Japan}
\affiliation[b]{Institute for Cosmic Ray Research, The University of Tokyo, 5-1-5 Kashiwanoha, Kashiwa, Chiba 277-8582, Japan}

\affiliation[c]{Institute of Cosmology \& Gravitation, University of Portsmouth, Dennis Sciama Building, Burnaby Road, 
Portsmouth, PO1 3FX, United Kingdom}

\emailAdd{yuichiro.tada@ipmu.jp}
\emailAdd{vincent.vennin@port.ac.uk}

\date{today}

\begin{document}
\sloppy

\abstract{The prospects of future galaxy surveys for non-Gaussianity measurements call for the development of robust techniques for computing the bispectrum of primordial cosmological perturbations. In this paper, we propose a novel approach to the calculation of the squeezed bispectrum in multiple-field inflation. With use of the $\delta N$ formalism, our framework sheds new light on the recently pointed out difference between the squeezed bispectrum for global observers and that for local observers, while allowing one to calculate both. For local observers in particular, the squeezed bispectrum is found to vanish in single-field inflation. Furthermore, our framework allows one to go beyond the near-equilateral (``small hierarchy'') limit, and to automatically include intrinsic non-Gaussianities that do not need to be calculated separately. The explicit computational programme of our method is given and illustrated with a few examples.
}



\begin{flushright}
IPMU 16-0143
\end{flushright}

\maketitle
\flushbottom

\section{Introduction}
\label{sec:intro}
Inflation~\cite{Starobinsky:1980te, Sato:1980yn, Guth:1980zm, Linde:1981mu, Albrecht:1982wi, Linde:1983gd} is the leading paradigm to describe the physical conditions that prevailed in the very early Universe. During this accelerated expansion epoch, vacuum quantum fluctuations of the gravitational and matter fields were amplified to large-scale cosmological perturbations~\cite{Starobinsky:1979ty, Mukhanov:1981xt, Hawking:1982cz,  Starobinsky:1982ee, Guth:1982ec, Bardeen:1983qw}, that later seeded the Cosmic Microwave Background (CMB) anisotropies and the large scale structure of our Universe. Inflation can proceed at energy scales as large as $10^{16}\,\GeV$ where particle physics remains elusive, which is why hundreds of scenarios have been proposed that implement inflation in different versions of high energy physics and gravity. However, cosmological observations, such as the recent Planck measurements~\cite{Ade:2013sjv, Adam:2015rua, Ade:2015xua, Ade:2015lrj} of the CMB temperature and polarisation anisotropies, have allowed one to start discriminating between these models~\cite{Martin:2013tda, Martin:2013nzq, Price:2015qqb, Vennin:2015eaa}. Single-field slow-roll models of inflation with a minimal kinetic term appear to be currently preferred, even if a large number of other scenarios still remain compatible with the data~\cite{Chen:2012ja, Vennin:2015vfa, Vennin:2015egh, Chen:2016vvw}.

A crucial observable to differentiate between these inflationary scenarios is the amount of non-Gaussianity (NG) they predict. This can be characterised by the non-linearity parameter $\fnl$, which measures the ratio between the bispectrum and the power spectrum squared. Current CMB measurements~\cite{Ade:2015ava} place the $68\%$ CL constraint $\fnl^\mathrm{local} = 0.8 \pm 5.0$ in the local configuration, $\fnl^\mathrm{equil}=-4\pm 43$ in the equilateral configuration and $\fnl^\mathrm{ortho}=-26\pm 21$ in the orthogonal configuration. In single-field slow-roll models of inflation, $\fnl$ is of the same order as the slow-roll parameters, that is to say $\sim 10^{-2}$, and these constraints are still too loose to use NG to discriminate between these scenarios. However, there are setups~\cite{Linde:1996gt, Enqvist:2001zp, Lyth:2001nq, Moroi:2001ct, Bartolo:2002vf, Dvali:2003ar, Elliston:2012wm} in which $\fnl$ can be made larger. In particular, if $\fnl$ is of order unity, it should be detectable by future galaxy surveys~\cite{Alvarez:2014vva} such as Euclid~\cite{Laureijs:2011gra}, SKA~\cite{Camera:2014bwa}, DESI~\cite{Levi:2013gra} and LSST~\cite{Abell:2009aa}. For such a low amplitude of the bispectrum signal, it seems therefore important to develop accurate methods for calculating $\fnl$, that notably account for the effect of hierarchies among the wavenumbers~\cite{Dias:2016rjq}.

If the fields contributing to the NG signal are all light during inflation, the squeezed limit of the bispectrum can be obtained through the $\delta N$ formalism~\cite{Starobinsky:1986fxa, Salopek:1990jq, Sasaki:1995aw, Sasaki:1998ug, Wands:2000dp, Lyth:2004gb, Lyth:2005fi}, which relates curvature perturbations to fluctuations in the number of \efolds realised along background trajectories. In its standard formulation, 
it allows one to recover the consistency relation (CR)~\cite{Maldacena:2002vr, Creminelli:2004yq} $\frac{3}{5}\fnl^\mathrm{CR} = \frac{1-\nS}{4}$ for the local configuration in the single-field case, where $\nS$ is the spectral index of the curvature perturbations power spectrum. The robustness of this approach in multiple-field inflation has also been successfully tested~\cite{Watanabe:2011sm}. However, it has recently been suggested~\cite{Urakawa:2009gb, Tanaka:2011aj, Pajer:2013ana, Bartolo:2015qva,Dai:2015jaa,dePutter:2015vga} 
that the squeezed bispectrum should vanish for a local observer in single-field slow-roll inflation, the CR being removed by properly fixing the physical wavelengths of the modes involved. 

In this paper, we provide new insight on this issue by reformulating the problem in field space. In the ``forward formulation'', the scales of the perturbations are defined  through the number of \efolds realised forwards between their Hubble exit times, and the results of the standard approach (including the CR) can easily be recovered. In the ``backward formulation'', the scales are defined with respect to a local observer at the end of inflation by evolving the equations of motion backwards. In this case, the squeezed bispectrum in single-field inflation is found to vanish exactly. Similarly to the methods recently proposed in \Refs{Kenton:2015lxa, Byrnes:2015dub, Kenton:2016abp} (that in fact correspond to our forward formulation), both the forward and backward formulations also have the benefit of automatically accounting for the ``intrinsic'' contribution to $\fnl$ (that comes from the intrinsic NG of the field fluctuations at Hubble crossing), which otherwise needs to be calculated separately from the third-order action in the standard procedure~\cite{Seery:2005gb}. In our approach, it arises as a simple slow-roll correction and the full result requires to integrate background dynamics only. Moreover, our framework accounts for arbitrary hierarchies among the wavenumbers involved in the bispectrum, \ie it implements the fact that different modes exit the Hubble radius at different times during inflation.

This paper is organised as follows. In \Sec{sec:GaugeArtifact}, we review definitions and properties related to the squeezed bispectrum of inflationary curvature perturbations, and recall why $\fnl^\mathrm{CR}$ is removed from $\fnl$ measured by a local observer.
In \Sec{sec:newApproach}, it is explained how $\fnl$ is usually calculated in the $\delta N$ formalism. 
Our new approach is presented, both in its forward and its backward formulation, 
and a schematic computational programme is given that explains how it can be implemented in practice. 
It is illustrated with two concrete examples in \Sec{sec:examples}, double massive inflation and inhomogeneous end of inflation, 
where the regions of parameter space where $\fnl$ is of order one or higher are identified. 
Finally, in \Sec{sec:conclusion}, we summarise our main results and draw a few concluding remarks.

\section{Local observer effect on the squeezed bispectrum}
\label{sec:GaugeArtifact}
Inflationary models can be compared on the basis of the statistics they predict for cosmological perturbations. The gauge-invariant curvature perturbation $\zeta$ is of particular interest, since it seeds the temperature fluctuations of the CMB and subsequently the density inhomogeneities of the Universe. Its two-point correlation function gives rise to the power spectrum $P_\zeta$, defined according to
\bea
	\braket{\zeta_{\bm{k}_1}\zeta_{\bm{k}_2}}=\int\dd^3x_1\dd^3x_2\,\ee^{-i(\bm{k}_1\cdot{\bm{x}}_1+\bm{k}_2\cdot{\bm{x}}_2)}
	\braket{\zeta({\bm{x}}_1)\zeta({\bm{x}}_2)}=(2\pi)^3\delta^{(3)}(\bm{k}_1+\bm{k}_2)P_\zeta(k_1).
\eea
In this expression, the momentum conservation term $(2\pi)^3\delta^{(3)}(\bm{k}_1+\bm{k}_2)$ implements the statistical homogeneity and isotropy condition, \ie the assumption that $\braket{\zeta({\bm{x}}_1)\zeta({\bm{x}}_2)}$ is a function of $|{\bm{x}}_1-{\bm{x}}_2|$ only. 
Equivalently, the power spectrum is also given by the direct Fourier transform of the two point function as
\bae{\label{eq: Pzeta as Fourier mode}
	P_\zeta(k)=\int\dd^3x\,\ee^{-i\bm{k}\cdot\bm{x}}\braket{\zeta(0)\zeta(\bm{x})}.
}
From $P_\zeta$, the dimensionless power spectrum 
\bea
\label{eq:calP:P}
\mathcal{P}_\zeta(k)=\frac{k^3}{2\pi^2}P_\zeta(k)
\eea
can also be defined. It is found to be almost scale invariant (\ie independent of $k$) in slow-roll inflation, and its departure from scale invariance is conveniently characterised by the spectral index
\bea
\label{eq:nS:def}
\nS = 1+ \frac{\dd\log \calP_\zeta}{\dd \log k},
\eea
observationally measured~\cite{Ade:2015lrj} to $\nS = 0.968 \pm 0.006$ at $68\%$ CL.
\subsection{Squeezed bispectrum}
Similarly to the power spectrum, the bispectrum $B_\zeta$ is obtained from the three-point correlation function according to
\bea
	\braket{\zeta_{\bm{k}_1}\zeta_{\bm{k}_2}\zeta_{\bm{k}_3}}=(2\pi)^3\delta^{(3)}(\bm{k}_1+\bm{k}_2+\bm{k}_3)
	B_\zeta\left(k_1,k_2,k_3\right).
\eea
If curvature perturbations are exactly Gaussian, the bispectrum vanishes, so that $B_\zeta$ is the lowest order correlation function characterising the non-Gaussian nature of $\zeta$.

As a typical NG profile, let us consider the ``local'' ansatz
\bea\label{local type}
	\zeta({\bm{x}})=g({\bm{x}})+\frac{3}{5}f_\mathrm{NL}^\mathrm{local}(g^2({\bm{x}})-\braket{g^2}),
\eea
where $g({\bm{x}})$ is a Gaussian field, and $\fnl^\mathrm{local}$ is called ``local'' non-linearity parameter. With this assumption, making use of Wick's theorem, the bispectrum is given by
\bea
\label{eq:bispectrum:local}
	B_\zeta(k_1,k_2,k_3)=\frac{6}{5} \fnl^{\mathrm{local}}\left[P_\zeta(k_1)P_\zeta(k_2)+P_\zeta(k_2)P_\zeta(k_3)+P_\zeta(k_3)P_\zeta(k_1)\right].
\eea
Let us now consider the squeezed limit where $k_\mathrm{L}=k_1\ll k_2\simeq k_3=k_\uS $. If $\zeta$ has an almost scale-invariant power spectrum, $P_\zeta(k)\sim k^{-3}$, 
the second terms in the brackets of \Eq{eq:bispectrum:local} is sub-dominant with respect to the two other ones, and one has
\bea
\label{squeezed bispectrum}
	B_\zeta(k_\mathrm{L},k_\uS ,k_\uS )\simeq\frac{12}{5}f_\mathrm{NL}^\mathrm{local}P_\zeta(k_\mathrm{L})P_\zeta(k_\uS ), 
	\quad \text{for $k_\mathrm{L}\ll k_\uS $}.
\eea 

More generally, the scale-dependent non-linearity parameter $\fnl$ for arbitrary NG types can be defined as
\bea
\frac{3}{5}\fnl \left(k_1,k_2,k_3\right) \equiv \frac{B_\zeta(k_1,k_2,k_3)}{2\left[P_\zeta(k_1)P_\zeta(k_2) + 2\,\mathrm{perms.}\,\right]},
\eea
where the two permutations are explicitly given in \Eq{eq:bispectrum:local}. In the squeezed limit, this gives rise to
\bea
\frac{3}{5}\fnl \left(k_\uL ,k_\uS \right) = \frac{B_\zeta(k_\uL ,k_\uS ,k_\uS )}{4P_\zeta(k_\uL )P_\zeta(k_\uS )}.
\eea

\subsection{Correlating long and short wavelength fluctuations}
\label{sec:correlatingShortLong}
From \Eq{squeezed bispectrum}, one can see that local-type NG represents a non-vanishing correlation between long and short wavelength fluctuations. This can be understood as follows. Let us consider a long and a short wavelength mode $k_\uL $ and $k_\uS $ in a given patch whose size $R$ satisfies $k_\uL ^{-1}\gg R\gg k_\uS ^{-1}$. In the local model of \Eq{local type}, one can expand $g({\bm{x}})$ into a long wavelength part $g_\mathrm{L}$ and a short wavelength part $g_\uS ({\bm{x}})$, and one obtains
\bea
	\zeta({\bm{x}})&=&g_\uL +\frac{3}{5}f_\mathrm{NL}^\mathrm{local}(g_\uL ^2-\braket{g^2})
\nonumber \\ &  &
	+g_\uS ({\bm{x}})+\frac{3}{5}f_\mathrm{NL}^\mathrm{local}\left[ g_\uS ^2({\bm{x}})+2g_\uL  g_\uS  ({\bm{x}})\right].
	\label{eq:zeta:shortlong}
\eea
Here, the ${\bm{x}}$-dependence of $g_\mathrm{L}$ is omitted since the long-wavelength mode is almost constant within the considered patch. Therefore, the second line of \Eq{eq:zeta:shortlong} represents the non-constant short-wavelength mode of $\zeta$, and at linear order in $g_\uS $, it reads
\bea
\label{modulation by gl} 
	\zeta_\uS ({\bm{x}})\sim \left(1+\frac{6}{5}f_\mathrm{NL}^\mathrm{local}g_\uL \right)g_\uS  ({\bm{x}}).
\eea
From this expression, one can see that a local-type NG yields a modulation of the amplitude of the short-wavelength mode by the long-wavelength mode.

From the particle physics perspective, correlations between long- and short-wavelength fluctuations are associated to soft-particle exchanges. Their properties therefore depend on the number and on the nature of the exchanged particles (invariant mass or spin)~\cite{Arkani-Hamed:2015bza}. If inflation is realised by a single slowly-rolling field, the squeezed $\fnl$ parameter can be calculated in the in-in formalism (see \Ref{Mukhanov:1990me} for a review) and one obtains the CR~\cite{Maldacena:2002vr, Creminelli:2004yq} mentioned in \Sec{sec:intro}
\bea
\label{CR}
\frac{3}{5}\fnl\left(k_\mathrm{L},k_\uS \right)=\frac{1-\nS}{4}.
\eea
From \Eq{modulation by gl}, this CR can naively be understood as the statement that  $\zeta_\uS $ is modulated by $\zeta_\uL $ according to $\zeta_\uS \to [1+(1-\nS)\zeta_\uL /2]\zeta_\uS $. With the definition~(\ref{eq:nS:def}) of the spectral index and schematically taking $\zeta_\uS \sim \sqrt{\calP_{\zeta}(k_\uS) }$, this modulation reads
\bea
	\Delta\zeta_\uS \sim\frac{1-\nS}{2}\zeta_\uL \zeta_\uS \sim-\dif{\log\zeta_\uS }{\log k_\uS}\zeta_\uL \zeta_\uS \sim-\dif{\zeta_\uS }{\log k_\uS}\zeta_\uL .
\eea
Interpreted as the chain rule of differentiation, this expression suggests that the effect of the long-wavelength modulation is the same as a simple scale shift 
\bea
\label{eq:scaleshift}
\log k_\uS \to\log k_\uS -\zeta_\uL.
\eea
\subsection{Squeezed bispectrum for a local observer}
\label{sec:projectionEffects}
These discussions imply that the CR~(\ref{CR}) can be viewed as a local rescaling of the physical wavelengths. 
Indeed, in the gauge where fixed $t$ slices of space-time have uniform energy density and fixed $\bm{x}$ worldlines are comoving, the curvature perturbation $\zeta$ can be defined as the spatial dependence of the scale factor $a$ on uniform density slices
\bea
\label{eq:a:zeta}
	a\left(t,{\bm{x}}\right)=a_0(t)\ee^{\zeta(t,{\bm{x}})}.
\eea
For a fixed comoving wavenumber $k$, the physical wavenumbers $k_\mathrm{ph}=k/a$ inside and outside the considered patch thus differ by $\Delta\log k_\mathrm{ph}=-\zeta_\mathrm{L}$, which exactly matches the scale shift~(\ref{eq:scaleshift}). Therefore, if scales are defined with respect to their physical wavenumbers, the correlations between long- and short-wavelengths discussed in \Sec{sec:correlatingShortLong} vanish.

For this reason, \Refs{Urakawa:2009gb,Tanaka:2011aj} recently pointed out that the squeezed bispectrum given by the CR can be removed by using the gauge degree of freedom associated to the finiteness of the observable universe. We call this effect the ``local observer effect''. The same conclusion was reached in \Ref{Pajer:2013ana} by introducing locally homogeneous isotropic coordinates called 
``conformal Fermi normal coordinates'', denoted $\FNC$. 
Let us briefly review their argument by considering again a patch of size $R$ satisfying $k_\uL ^{-1}\gg R\gg k_\uS ^{-1}$, 
at a time when the long-wavelength mode is super-Hubble, $k_\mathrm{L}\ll aH$, and the short-wavelength mode is 
sub-Hubble or of the order of the Hubble scale, $k_\uS \gtsim aH$. 
Including long-wavelength scalar perturbations only, in the same gauge that was employed in \Eq{eq:a:zeta}, the metric is given by
\bea
	\dd s^2=-\dd t^2+a^2(t)\left[1+2\zeta_\mathrm{L}({\bm{x}})\right]\dd{\bm{x}}^2+\mathcal{O}\left(\frac{k_\mathrm{L}^2}{a^2H^2}\right).
\eea
The $\FNC$ are then defined according to
\bea
\label{eq:fnc:def}
	\overline{{\bm{x}}}_\mathrm{F}=\left[1+\zeta_\mathrm{L}({\bm{x}}=0)\right]{\bm{x}}.
\eea
Since the long-wavelength mode is almost constant in the patch we consider, the metric can be expanded as
\bea
	\dd s^2=-\dd t^2+a^2(t)\dd\overline{{\bm{x}}}_\mathrm{F}^2+\mathcal{O}(k_\mathrm{L}^2\overline{x}_\mathrm{F}^2)+\mathcal{O}\left(\frac{k_\mathrm{L}^2}{a^2H^2}\right),
\eea
where $\overline{x}_\mathrm{F}\ll k_\mathrm{L}^{-1}$ inside the patch. The long-wavelength metric perturbation therefore disappears in the $\FNC$. Let us note that this does not mean that the gauge-invariant curvature perturbation $\zeta$ vanishes, since it simply transforms as a scalar, $\overline{\zeta}\left[\overline{x}_\mathrm{F}(x)\right]=\zeta(x)$. This is why, at linear order, the two-point function of $\zeta_\uS $ transforms according to
\bea
\label{eq:fnc:2pt:trans}
	\braket{\overline{\zeta}_\uS (\overline{{\bm{x}}}_\mathrm{F})\overline{\zeta} _\uS (0)}
	=\braket{\zeta_\uS \left[{\bm{x}}(\overline{{\bm{x}}}_\mathrm{F})\right]\zeta_\uS (0)} 
	\simeq \left[1 - \zeta_\uL (0)\overline{x}_{\mathrm{F}i}\pdif{}{\overline{x}_{\mathrm{F}i}}\right] \braket{\zeta_\uS (\overline{{\bm{x}}}_\mathrm{F})\zeta_\uS (0)},
\eea
where $i$ is implicitly summed over. Configurations mostly contributing to the squeezed bispectrum are such that $|{\bm{x}}_1-{\bm{x}}_2|\gg|{\bm{x}}_2-{\bm{x}}_3|$, where only the long-wavelength mode can contribute to $\zeta({\bm{x}}_1)$. The squeezed three-point function $\braket{\zeta({\bm{x}}_1)\zeta({\bm{x}}_2)\zeta({\bm{x}}_3)}$ can then be understood as the modulation of the small scale two-point correlator under to the long-wavelength mode, $\braket{\zeta_\uL ({\bm{x}}_1)  \braket{ \zeta_\uS ({\bm{x}}_2)\zeta_\uS ({\bm{x}}_3) }}$. Making use of the transformation rule~(\ref{eq:fnc:2pt:trans}), the squeezed bispectrum then transforms according to
\bea
\label{eq:fnc:squeezed3pt:trans}
	B_{\overline{\zeta}}(k_\mathrm{L},k_\uS ,k_\uS ) &= &B_\zeta(k_\uL ,k_\uS ,k_\uS )+P_\zeta(k_\uL )\partial_{k_{\uS ,i}}\left[k_{\uS ,i}P_\zeta(k_\uS )\right] \nonumber \\
	&=& B_\zeta(k_\uL ,k_\uS ,k_\uS )+\left[3+\dif{\log P_\zeta(k_\uS )}{\log k_\uS }\right]P_\zeta(k_\uL )P_\zeta(k_\uS ),
\eea
where integration by parts has been performed and the relation $\partial_i(x_i\ee^{-i\bm{k}\cdot{\bm{x}}}) =\partial_{k_i}(k_i\ee^{-i\bm{k}\cdot{\bm{x}}})$ has been used. Combining \Eqs{eq:calP:P} and~(\ref{eq:nS:def}), one can see that the terms inside the brackets of the second line of \Eq{eq:fnc:squeezed3pt:trans} reduce to $\nS(k_\uS)-1$, so that if the original bispectrum is given by the CR (\ref{CR}), the bispectrum calculated in the $\FNC$ vanishes when expressed in terms of physical scales.

At this point, it is worth stressing~\cite{Bartolo:2015qva,Dai:2015jaa,dePutter:2015vga} 
that even though the $\FNC$ leads to a vanishing CR in local patches, \Eq{eq:fnc:def} is only defined locally and there is no global gauge transformation that can eliminate the bispectrum of the gauge-invariant curvature perturbations. However, in single-field inflation, the only effect of the inflaton perturbations is to swing the evolution progress back and forth along the attractor trajectory (which is why such perturbations are called ``adiabatic''). Each Hubble patch thus emerges with the same initial conditions for the short-wavelength modes, the only differences being related to changes in the evolution progress, 
and the correlation between the long- and short-wavelength modes vanishes in the local patch.

Let us also note that in practice, an apparent bispectrum re-appears in concrete observables, 
such as the CMB angular correlation functions or the scale-dependent bias, 
due to projection effects~\cite{Pajer:2013ana, Bartolo:2015qva,Dai:2015jaa,dePutter:2015vga}. These are related to the fact that since the long-wavelength mode is observed inside our horizon, it affects the mapping of the actual positions of the light sources on the celestial sphere. However, these projection effects can be evaluated separately, once the original bispectrum is calculated in terms of physical scales.

In the following, we present a new approach to the calculation of the squeezed bispectrum in the $\delta N$ formalism. This allows us to reformulate the difference between the squeezed bispectrum of $\zeta$ (``forward'' formulation, see \Sec{sec:forwardformulation}), given by the CR~(\ref{CR}) in the single-field case, 
and the one of $\bar{\zeta}$ (``backward'' formulation, see \Sec{sec:backward formulation}), which vanishes in single-field inflation. 
\section{Squeezed bispectrum in the $\delta N$ formalism}
\label{sec:newApproach}
Let us consider the case of one or several scalar fields minimally coupled to gravity,
\bea
	S=\int\dd^4x\left[\frac{1}{2}\Mp^2R-\frac{1}{2}\mathcal{G}_{ij}\partial_\mu\phi^i\partial^\mu\phi^j-V(\phi)\right].
\eea
Hereafter, $\Mp=\sqrt{8\pi G}^{-1}\simeq 2.4\times10^{18}\,\mathrm{GeV}$ denotes the reduced Planck mass, 
and roman indices label the different scalar fields. The scalar potential $V(\phi)$ is assumed to support a phase of slow-roll inflation, 
and all calculations will be performed at leading order in slow roll, though our formalism can be extended to non slow-roll dynamics as long as
the system evolves along a phase-space attractor. 
Unless stated otherwise, the field space metric $\mathcal{G}_{ij}$ is assumed to be almost flat for the field values of interest 
so that it can be reabsorbed in  canonical renormalisations of the fields under which $\mathcal{G}_{ij}\to\delta_{ij}$. 
Along the slow-roll attractors, the fields evolve according to
\bea
\label{eq:SRtraj}
\frac{\dd\phi_i}{\dd N} = -\Mp^2 \frac{V_i}{V}.
\eea

In this section, we first review the usual method to calculate the squeezed bispectrum in the $\delta N$ formalism, 
which can only be applied to near-equilateral configurations as will be explained. 
We then introduce an alternative algorithm that goes beyond near-equilateral configurations, and that can be applied to calculating correlations between 
physical scales in local patches, as discussed in \Sec{sec:projectionEffects}.
\subsection{Standard approach}
\label{sec:standardApproach}
As mentioned in \Sec{sec:intro}, in the $\delta N$ formalism~\cite{Starobinsky:1986fxa, Salopek:1990jq, Sasaki:1995aw, Sasaki:1998ug, Wands:2000dp, Lyth:2004gb, Lyth:2005fi}, the super-Hubble curvature perturbations are given by the spatial differences in the number of \efolds realised between an initial flat hypersurface and a final uniform density hypersurface. Such fluctuations of the number of \efolds are denoted $\delta N$ and usually approximated by a perturbative expansion around the background field values
\bea\label{delta N expansion}
	\zeta({\bm{x}})=\delta N({\bm{x}})=N_i(\phi_*)\delta\phi^i({\bm{x}})
	+\frac{1}{2}N_{ij}(\phi_*)\delta\phi^i({\bm{x}})\delta\phi^j({\bm{x}})+\cdots.
\eea 
In this expression, $N(\phi_*)$ denotes the number of \efolds realised from the initial field value $\phi_*$ and until a given final uniform density hypersurface is reached, and $N_i=\partial N/\partial\phi_*^i$ and $N_{ij}=\partial^2N/(\partial\phi_*^i\partial\phi_*^j)$ are its derivatives with respect to the field values $\phi_*$. From here, at leading order in perturbation theory, the power spectrum of curvature perturbations can be expressed as\footnote{In the case of a general field space metric, the Kronecker $\delta^{ij}$ should simply be replaced by the field metric $\mathcal{G}^{ij}$.}
\bea
\label{eq:Pzeta:deltaN}
	\mathcal{P}_\zeta(k)=N_iN_j\frac{k^3}{2\pi^2}\int\dd^3x\,\ee^{-i\bm{k}\cdot\bm{x}}\braket{\delta\phi^i(0)\delta\phi^j(\bm{x})}
	=N_iN_j\delta^{ij}\mathcal{P}_{\delta\phi},
\eea
where $\mathcal{P}_{\delta \phi}$ denotes the power spectrum of the field fluctuations at Hubble exit time, given by $\mathcal{P}_{\delta\phi} = \left(\frac{H}{2\pi}\right)^2$ at leading order in slow roll.

A similar expression can be obtained for the bispectrum of curvature perturbations, 
\bea
	B_\zeta(k_1,k_2,k_3)=N_iN_jN_kB_{\phi^i\phi^j\phi^k}(k_1,k_2,k_3)+\left[N_iN_jN_{ij}P_\phi(k_1)P_\phi(k_2) + \text{2 perms}\right].
\eea
In this expression, the first term comes from the NG of the scalar field fluctuations themselves and is called ``intrinsic" NG, while the second term is due to the higher order expansion in $\delta N$ and is called ``$\delta N$ component'' of the bispectrum. If one parametrises the intrinsic NG as
\bea	
	B_{\phi^i\phi^j\phi^k}=\frac{\left(2\pi^2 \calP_{\delta\phi}\right)^2}{\left(k_1k_2k_3\right)^3}\mathcal{A}^{ijk}(k_1,k_2,k_3),
\eea
the non-linearity parameters corresponding to the intrinsic and $\delta N$ NG are respectively given by
\begin{align}
	\frac{3}{5}\fnl^\mathrm{int}=\frac{\mathcal{A}^{ijk}(k_1 , k_2 , k_3 )N_iN_jN_k}{2(N_lN_l)^2\left(k_1^3+k_2^3+k_3^3\right)}, \quad 
	\frac{3}{5}\fnl^{\delta N}=\frac{N_iN_jN_{ij}}{2(N_kN_k)^2},
\end{align}
where $\fnl = \fnl^\mathrm{int} + \fnl^{\delta N}$. In \Ref{Seery:2005gb}, it is explained how the quantity $\mathcal{A}^{ijk}$ can be calculated from the third-order action, and one obtains
\bea
	\mathcal{A}^{ijk}(k_1,k_2,k_3)=\sum_\text{6 perms}\frac{V_i}{4V}\delta^{jk}\left[3\frac{k_2^2k_3^2}{k_{\mathrm{t}}}+\frac{k_2^2k_3^2}{k_{\mathrm{t}}^2}(k_1+2k_3)-\frac{1}{2}k_1^3+k_1k_2^2\right], 
\eea
where $k_{\mathrm{t}}=k_1+k_2+k_3$. 
In this expression, the sum is over the six permutations of $(ijk)$, where the momenta $k_1$, $k_2$, and $k_3$ are also rearranged accordingly.  
In the squeezed limit, taking \eg $k_\uL =k_1\ll k_2\simeq k_3=k_\uS $, this expression boils down to
\bea
	\mathcal{A}^{ijk}=k_\uS ^3\frac{V_i}{V}\delta^{jk},
\eea
and the non-linearity parameter reads~\cite{Vernizzi:2006ve,Yokoyama:2007uu,Yokoyama:2007dw}
\bea
\label{fNL std}
	\frac{3}{5}f_\mathrm{NL}=\frac{3}{5}f_\mathrm{NL}^\mathrm{int}+\frac{3}{5}f_\mathrm{NL}^{\delta N}
	=\dfrac{N_i\dfrac{V_i}{V}}{4N_jN_j}+\dfrac{N_iN_jN_{ij}}{2(N_kN_k)^2}.
\eea
From this expression, the single-field CR can easily be recovered. In this case indeed, one has $N_\phi = 1/(\sqrt{2\epsilon_1}\Mp)$, $V_\phi/V = \sqrt{2\epsilon_1}/\Mp$ and $N_{\phi\phi} = \epsilon_2/(4\epsilon_1\Mp^2)$, where $\epsilon_1$ and $\epsilon_2$ are the two first Hubble-flow parameters~\cite{Liddle:1994dx}. This gives rise to $\frac{3}{5}\fnl^{\mathrm{int}} = \frac{\epsilon_1}{2}$ and $\frac{3}{5}\fnl^{\delta N} = \frac{\epsilon_2}{4}$, hence
\bea
	\frac{3}{5}f_\mathrm{NL}=\frac{2\epsilon_1+\epsilon_2}{4}=\frac{1-\nS}{4}, 
\eea
\ie the CR~(\ref{CR}). In the previous expressions, one should note that the different quantities are evaluated at the same time. 
This is why, strictly speaking, they are valid in the near-equilateral configuration $k_1\ltsim k_2\sim k_3$ only. Since this breaks the hierarchy between the scales $k_\uL$ and $k_\uS$ on which the calculation of the squeezed configuration relies, this ``soft limit'' is not guaranteed to be always consistent, and in the following we go beyond this approximation.
\subsection{New approach}
\label{sec:alternative} 
Let us now see how the previous result can be extended to arbitrary separations between the scales $k_\uL$ and $k_\uS$. As discussed in \Sec{sec:GaugeArtifact}, the squeezed bispectrum is related to the correlation between the long-wavelength perturbation and the short-wavelength two-point function. 
We consider a patch of comoving size $k_\uL ^{-1}$, where $\zeta_\uL $ is the coarse-grained curvature perturbation on this scale defined through the window function $W$,
\bae{
	\zeta_\uL(\bm{x})=\int\dd^3yW(k_\uL|\bm{x}-\bm{y}|)\zeta(\bm{y}).
}
If the power spectrum $\mathcal{P}_\zeta(k_\uS )$ is evaluated within this local patch, one has
the following expressions, making use of Eqs.~(\ref{eq: Pzeta as Fourier mode}) and (\ref{eq:calP:P}).
\bea\label{zetaLcalPzetaS}
	\braket{\zeta_\uL \mathcal{P}_\zeta(k_\uS )}&=&\dfrac{k_\uS ^3}{2\pi^2}\int\dd^3x\,\dd^3y\,\ee^{-i\bm{k}_\uS \cdot\bm{y}}W(k_\uL x)
	\left\langle{\zeta({\bm{x}})\zeta\left(-\frac{\bm{y}}{2}\right)\zeta\left(\frac{\bm{y}}{2}\right)}\right\rangle \nonumber \\
	&=&\dfrac{k_\uS ^3}{2\pi^2}\int\dd^3x\,\dd^3y\int\frac{\dd^3p\,\dd^3q}{(2\pi)^6}
	\exp\left\lbrace i\left[\bm{p}\cdot\left({\bm{x}}+\frac{\bm{y}}{2}\right)+\bm{q}\cdot\bm{y}-\bm{k}_\uS \cdot\bm{y}\right]\right\rbrace
	W(k_\uL x)B_\zeta(p,q,|\bm{p}+\bm{q}|) \nonumber \\
	&=&\dfrac{k_\uS ^3}{2\pi^2}\int\frac{\dd^3p\,\dd^3q}{(2\pi)^3}\delta^{(3)}\left(\frac{\bm{p}}{2}+\bm{q}-\bm{k}_\uS \right)\tilde{W}\left(\frac{p}{k_\uL} \right)
	B_\zeta(p,q,|\bm{p}+\bm{q}|).
\eea
In this expression, since the Fourier transform of the window function $\tilde{W}(p/k_\uL )$ selects out the modes such that $p\ltsim k_\uL$, the delta function and the bispectrum can be approximated by $\delta^{(3)}({\bm{p}}/{2}+\bm{q}-\bm{k}_\uS )\simeq\delta^{(3)}(\bm{q}-\bm{k}_\uS )$ and  $B_\zeta(p,q,|\bm{p}+\bm{q}|)\simeq B_\zeta(p,k_\uS ,k_\uS )$, and one obtains
\bea
\label{eq:zetaLPzeta:corr}
	\braket{\zeta_\uL \mathcal{P}_\zeta(k_\uS )}\simeq\frac{k_\uS ^3}{2\pi^2}\int^{\log k_\uL }\frac{p^3}{2\pi^2}B_\zeta(p,k_\uS ,k_\uS )\dd\log p.
\eea
The squeezed non-linearity parameter is then given by
\bea
	\frac{3}{5}f_\mathrm{NL}(k_\uL ,k_\uS )=\frac{1}{4\mathcal{P}_\zeta(k_\uL )\mathcal{P}_\zeta(k_\uS )}\dif{\braket{\zeta_\uL \mathcal{P}_\zeta(k_\uS )}}{\log k_\uL }.
\eea

In this expression, since $\zeta_\uL$ includes all fluctuations with wavelengths larger than $k_\uL^{-1}$, evaluating the bispectrum when its first argument is $k_\uL$ requires to differentiate the integral in \Eq{eq:zetaLPzeta:corr} with respect to $\log k_\uL$. In the $\delta N$ formalism, a different method is commonly used, which relies on the introduction of a single-mode impulsive field fluctuation
\bea
	\delta\phi_\uL(\bm{x})=\int_{\log p=\log k_\uL}\frac{\dd^3p}{(2\pi)^3}\ee^{i\bm{p}\cdot\bm{x}}\delta\phi_{\bm{p}},
\eea
yielding a single-mode number of \efolds fluctuation
\bea\label{eq: delta NL}
	\delta N_\uL(\bm{x})=\int_{\log p=\log k_\uL}\frac{\dd^3p}{(2\pi)^3}\ee^{i\bm{p}\cdot\bm{x}}\zeta_{\bm{p}}.
\eea
The correlation $\braket{\delta N_\uL\calP_\zeta(k_\uS)}$ thus includes only the modes $k_\uL$ and $k_\uS$
and allows one to avoid differentiation with respect to $\log k_\uL$,\footnote
{In this case, the window function $\tilde{W}$ in \Eq{zetaLcalPzetaS}, typically given by a step function $\theta(\log p-\log k_\mathrm{L})$, should be replaced by a Dirac function $\delta(\log p-\log k_\mathrm{L})$.
\label{footnote:windowF}}
\bea
\label{fNL w/ impulsive fluctuation}
	\frac{3}{5}\fnl \left(k_\uL ,k_\uS \right)=\frac{\braket{\delta N_\uL \mathcal{P}_\zeta(k_\uS )}}{4\mathcal{P}_\zeta(k_\uL )\mathcal{P}_\zeta(k_\uS )}.
\eea
\subsubsection{Forward formulation}
\label{sec:forwardformulation}
\begin{figure}[t]
\begin{center}
\includegraphics[width=0.7\textwidth]{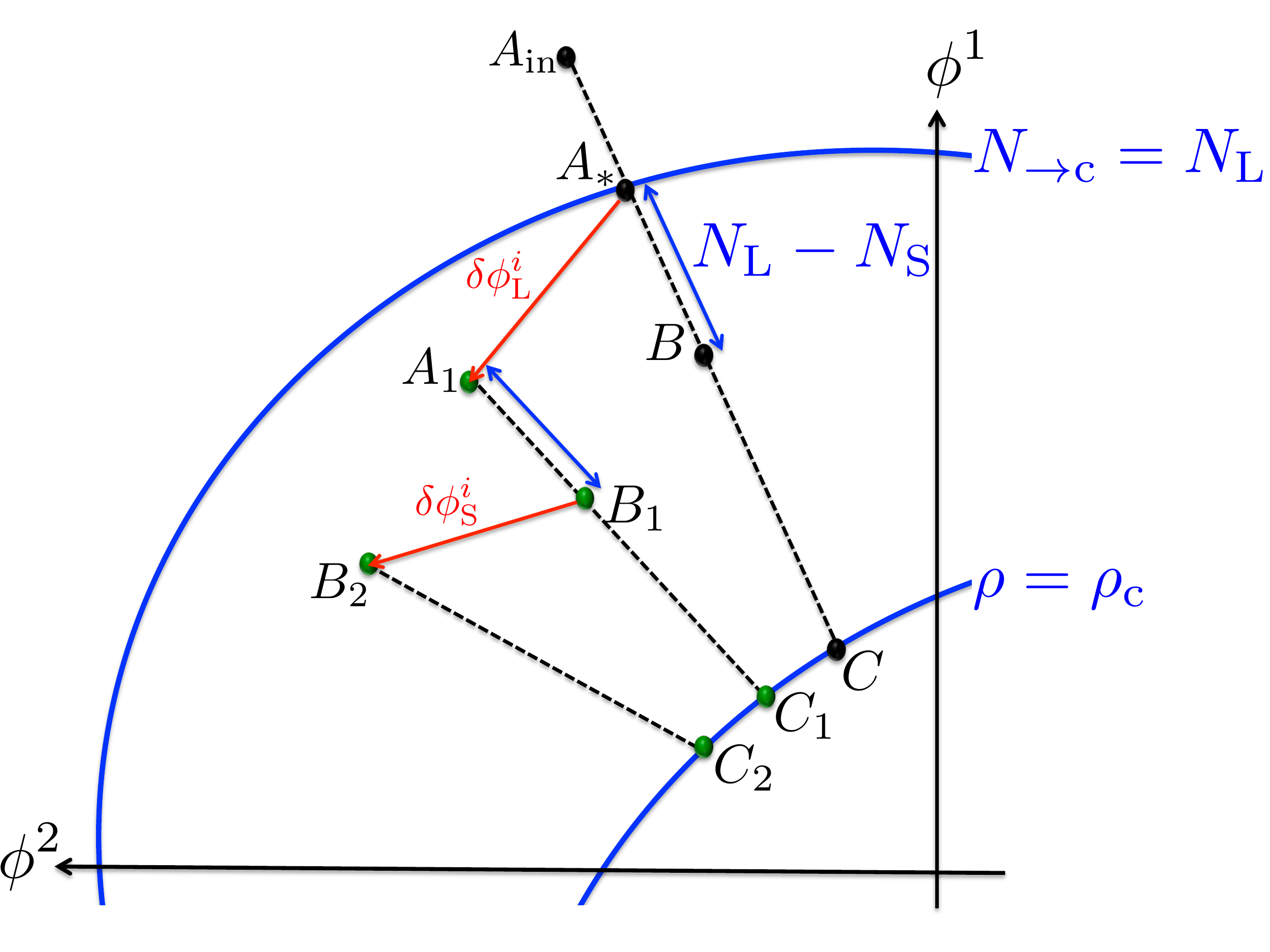}
\caption{Schematic representation of the forward procedure for the calculation of the squeezed bispectrum in the $\delta N$ formalism. In this formulation, the forward number of \efolds realised between $A_1$ and $B_1$ is fixed, $N_{A_1 B_1} = N_\uL - N_\uS$.}
\label{fig:schemaclass}
\end{center}
\end{figure}
Let us now see how \Eq{fNL w/ impulsive fluctuation} can be evaluated in practice in the $\delta N$ formalism, and how the results from \Refs{Kenton:2015lxa, Byrnes:2015dub, Kenton:2016abp} can be rederived. 
In \Fig{fig:schemaclass}, we describe how this is done in the forward formulation. The black dotted lines represent slow-roll attractor trajectories, $[A_\mathrm{in} C]$ being the unperturbed trajectory. The point $A_*$ is where the long-wavelength mode emerges from the Hubble radius. At this point, the field fluctuation $\delta\phi_\uL ^i$ leads to the variation $\delta N_\uL =N(A_1)-N(A_*)$ in the number of $e$-folds, that can be identified with the long-wavelength curvature perturbation $\zeta_\uL $. This fluctuation also shifts the field space trajectory to a different slow-roll solution, $[A_1 C_1]$, along which the short-wavelength perturbation emerges at $B_1$. Since the location of $B_1$ depends on $\delta\phi_\uL ^i$, the long-wavelength curvature perturbation $\zeta_\uL $ and the short-wavelength power spectrum $\calP_\zeta(B_1)$ are correlated
\bea
\label{dNLcalPS}
	\braket{\delta N_\uL \mathcal{P}_\zeta(k_\uS )}\simeq N_i(A_*)\left.\pdif{\mathcal{P}_\zeta}{\phi^j}\right|_{B}
	\braket{\delta\phi_\uL ^i\delta\tilde{\phi}_\uL ^j}.
\eea
In this expression, $\delta\tilde{\phi}_\uL ^i$ denotes the value of $\delta \phi_\uL ^i$ evolved after $N_\uL -N_\uS $ $e$-folds, that is, $\delta\tilde{\phi}_\uL ^i=\phi^i(B_1)-\phi^i(B)$. 
At leading order, it can be expressed as $\delta\tilde{\phi}_\uL^i=\partial\phi_B^j/\partial\phi^k|_*\delta\phi_\uL^k$ where $\partial\phi_B^j/\partial\phi^k|_*$ encodes the variation of the coordinates of $B$
due to the field fluctuations $\delta\phi_\uL$ at $A_*$. Then, with use of the field fluctuations power spectrum $\braket{\delta\phi_\uL^i\delta\phi_\uL^j}=\delta^{ij}\mathcal{P}_{\delta\phi}$, 
one obtains\footnote{\label{footnote:Gamma}With use of 
the so-called $\Gamma$ expansion technique~\cite{Yokoyama:2007uu, Yokoyama:2007dw, Seery:2012vj, Kenton:2015lxa, Byrnes:2015dub, Kenton:2016abp}, one can show that this relation is consistent with the results of 
\Ref{Kenton:2015lxa, Kenton:2016abp}. Let us note indeed that the fluctuation in the number of \efolds between the initial flat hypersurface and the final uniform density hypersurface does not depend on the time at which the flat hypersurface is initiated~\cite{Lyth:2004gb}. Therefore, the quantity $N_i\delta\phi_{\uL}^i$ can be evaluated at anytime after Hubble exit of the long-wavelength perturbation, and in particular, one has $N_i(A_*)\delta\phi_\uL ^i=N_i(B)\delta\tilde{\phi}_\uL ^i$. This is why \Eq{dNLcalPS} can be expressed as
\bea
	\braket{\delta N_\uL \mathcal{P}_\zeta(k_\uS )}=N_i(B)\left.\pdif{\mathcal{P}_\zeta(k_\uS )}{\phi^j}\right|_{B}
	\left\langle {\delta\tilde{\phi}_\uL ^i\delta\tilde{\phi}_\uL ^j} \right\rangle
	=N_i(B)\left.\pdif{\mathcal{P}_\zeta(k_\uS )}{\phi^j}\right|_B\Gamma_{ik}^{(B,*)}\Gamma_{jk}^{(B,*)}\left(\frac{H}{2\pi}\right)^2,
\eea
where $\Gamma_{ij}^{(B,*)}\equiv \left.\pdif{\phi_B^i}{\phi^j}\right|_*$ encodes the variations in the coordinates of $B$ under the field fluctuations $\delta\phi_\uL$ at $A_*$. Since $\mathcal{P}_\zeta=N_iN_i\left(\frac{H}{2\pi}\right)^2$, this leads to the non-linearity parameter
\bea
\label{Gamma expansion}
	\frac{3}{5}f_\mathrm{NL}=\dfrac{N_i(B)N_l(B)N_{jl}(B)\Gamma_{ik}^{(B,*)}\Gamma_{jk}^{(B,*)}}{2N_m(A_*)N_m(A_*)N_n(B)N_n(B)}
	+\dfrac{N_i(B)\left.\dfrac{V_j}{V}\right|_B\Gamma_{ik}^{(B,*)}\Gamma_{jk}^{(B,*)}}{4N_m(A_*)N_m(A_*)}.
\eea
Making use of the relation $N_i(A_*)=N_j(B)\Gamma_{ji}^{(B,*)}$ and of \Eq{eq:SRtraj}, one indeed recovers Eq.~(3.3.3) of \Ref{Kenton:2015lxa} or Eq.~(4.8) of \Ref{Kenton:2016abp}.
}
\begin{align}\label{fNL beyond equilateral}
	\frac{3}{5}f_\mathrm{NL}=\frac{\braket{\delta N_\uL\mathcal{P}_\zeta(k_\uS)}}{4\mathcal{P}_\zeta|_*\mathcal{P}_\zeta|_B}
	=\frac{N_i|_*\left.\pdif{\mathcal{P}_\zeta}{\phi^j}\right|_B\left.\pdif{\phi^j_B}{\phi^i}\right|_*\mathcal{P}_{\delta\phi}|_*}{4\mathcal{P}_\zeta|_*\mathcal{P}_\zeta|_B}.
\end{align}
In order to calculate the power spectrum at small scales, an impulsive field fluctuation $\delta\phi_\uS ^i$ can be added from $B_1$ and the difference in the number of \efolds $\delta N_\uS =N(B_1)-N(B_2)$ yields the small-scale power spectrum. Let us also note that in this framework, the \emph{forward}\footnote{Note that, to calculate $\delta N$ itself, the backward number of \efolds function $N(\phi)$ is employed, 
but the forward number of \efolds is used when relating a perturbation scale with the location in field space where it exits the Hubble radius.} 
number of \efolds $N_\uL -N_\uS $ is used to determine the location of the point $B_1$ where small-scale fluctuations emerge, hence the name of the formulation.

This indeed seems natural in the context of the $\delta N$ formalism since it implies that, if large-scale fluctuations $\delta\phi_\uL ^i$ are defined on a spatially flat hypersurface, evolving this hypersurface by a uniform $N_\uL -N_\uS $ number of \efolds conserves its flatness. This is why in the forward formulation, scales can be uniformly defined in comoving coordinates, which however does not imply that they lead to the same physical scales on the uniform density hypersurface. If the physical scales on the uniform density hypersurface are used instead, one obtains the backward formulation that will be discussed in \Sec{sec:backward formulation}.\\

\noindent\textbf{Single-field case}\\
For now, let us show that the CR~(\ref{CR}) can properly be recovered in single-field inflation. In this case, the background field value $\phi$ and the backward number of \efolds $N$ have a one-to-one correspondence which is why one can label field space with $N$ instead of $\phi$. Making use of the relation $\delta N_\uL=N_\phi(A_*)\delta\phi_\uL=N_\phi(B)\delta\tilde{\phi}_\uL$ obtained
in footnote~{\ref{footnote:Gamma}}, \Eq{dNLcalPS} then gives rise to
\bea
\label{eq:forward:SF:deltaNlPzetaS}
	\braket{\delta N_\uL \mathcal{P}_\zeta(k_\uS )}\simeq\left.\pdif{\mathcal{P}_\zeta}{N}\right|_{B}\braket{\delta N_\uL ^2}.
\eea
In this expression, $\braket{\delta N_\uL ^2}$ is nothing but the power spectrum $\mathcal{P}_\zeta(k_\uL )$ since $\delta N_\uL$ only incorporates $k_\uL$ fluctuations (see Eq.~(\ref{eq: delta NL}) and 
footnote~\ref{footnote:windowF}), 
and the derivative of the power spectrum with respect to $N$ yields, by definition, the spectral index
\bea
	\left.\pdif{\mathcal{P}_\zeta}{N}\right|_{B}=\left.-\mathcal{P}_\zeta\pdif{\log\mathcal{P}_\zeta}{\log k}\right|_{B}
	=\left.\left(1-\nS\right)\mathcal{P}_\zeta\right|_{B},
\eea
see \Eq{eq:nS:def}. Therefore, the non-linearity parameter~(\ref{fNL w/ impulsive fluctuation}) is given by the CR 
\bea
\label{eq:forward:SF:fnl}
	\frac{3}{5}\fnl\left(k_\uL ,k_\uS\right)=\frac{1-\nS\left(k_\uS\right)}{4}.\eea\\

\noindent\textbf{Near-equilateral limit}\\
Let us also notice that the results of the standard approach presented in \Sec{sec:standardApproach} can be obtained from the forward formulation in the near-equilateral limit $N_\uS \to N_\uL$.
In this regime indeed, $B\to A_*$ and \Eq{fNL beyond equilateral} directly gives rise to \Eq{fNL std}. Let us however note that in the alternative approach presented here, there is no need to calculate $\fnl^\mathrm{int}$ separately from the cubic action and that this term is already incorporated in the $\delta N$ formalism.\footnote
{Note that here, the near-equilateral limit does not refer to the non-linearity parameter in the equilateral configuration $f_\mathrm{NL}^\mathrm{equil}$, but to the squeezed non-linearity parameter in the limit where the hierarchy between the two scales $k_\uL$ and $k_\uS$ can be neglected. Even though it may provide useful order of magnitude estimates, as already stressed at the end of \Sec{sec:standardApproach} this is a priori not fully consistent since the squeezed limit precisely relies on this hierarchy.}  

The interpretation of the two contributions $\fnl^\mathrm{int}$ and $\fnl^{\delta N}$ also become clearer. 
Since $\calP_\zeta = N_i N_i \mathcal{P}_{\delta\phi}$, see \Eq{eq:Pzeta:deltaN}, the derivatives of the power spectrum with respect the field values appearing in \Eq{dNLcalPS} contain two terms: one proportional to the derivative of $N_i$, that yields $\fnl^{\delta N}$, and one proportional to the derivative of $\calP_{\delta\phi}$, that yields $\fnl^\mathrm{int}$. Therefore, the so-called ``intrinsic'' NG is nothing but the effect of the field dependence of the amplitude of the field fluctuations.\\

\noindent\textbf{Non-canonical kinetic term case}\\
Let us finally see how the forward formulation can be extended to non-canonical kinetic terms. More precisely, we consider the case of single-field k-inflation~\cite{ArmendarizPicon:1999rj}
\bea
	S=\int\dd^4x\sqrt{-g}\left[\frac{\Mp^2}{2}R+P(X,\phi)\right],
\eea
where $X=-g^{\mu\nu}\partial_\mu\phi\partial_\nu\phi/2$. 
Here, we assume that the system has reached the phase-space attractor, along which the scalar field $\phi$ is not necessarily slowly rolling in k-inflation.
In \Refs{Gong:2015ypa, Domenech:2016zxn}, the $\delta N$ and intrinsic components of the non-linearity parameter are calculated and one has
\bea
\label{k-inflation}
	\frac{3}{5}\fnl^{\delta N}=\frac{1}{2}(\epsilon_1+\delta), \quad		
	\frac{3}{5}\fnl^\mathrm{int}=\frac{1}{4}\left(\epsilon_2-2\delta+s\right),
\eea
where $\epsilon_1$ and $\epsilon_2$ are the two first Hubble-flow parameters as before, $\delta\equiv \ddot{\phi}/(H \dot{\phi})$, $s\equiv \dotcs/(H\cs)$ and $\cs$ is the sound speed $c_s^{-2} \equiv 1+2XP_{XX}/P_X$. By summing up these two components, one obtains the CR~(\ref{CR})
\bea
\label{k-inflation:fnltotal}
	\frac{3}{5}\fnl=\frac{3}{5}\fnl^{\delta N}+\frac{3}{5}\fnl^\mathrm{int}=\frac{1}{4}\left(2\epsilon_1+\epsilon_2+s\right)=\frac{1-\nS}{4}.
\eea
These formulas can be recovered as follows. The $\delta N$ component of the non-linearity parameter is the same as in the standard case, and as mentioned above, the intrinsic component is related to the field derivative of $\calP_{\delta\phi}$, which in the near-equilateral limit $N_\uS \to N_\uL $ is given by
\bea
	\frac{3}{5}f_\mathrm{NL}^\mathrm{int}=\frac{N_\phi N_\phi^2\partial_\phi(\mathcal{P}_{\delta\phi})\mathcal{P}_{\delta\phi}}{4\mathcal{P}_\zeta^2}
	=\frac{\partial_\phi\log\mathcal{P}_{\delta \phi}}{4N_\phi}.
\eea
In k-inflation, the power spectrum of $\delta\phi$ reads~\cite{Garriga:1999vw}
\bea
	\mathcal{P}_{\delta \phi}=\frac{\dot{\phi}^2}{H^2}\mathcal{P}_\zeta=\frac{\dot{\phi}^2}{H^2}\frac{1}{2\epsilon_1 c_s\Mp^2}\left(\frac{H}{2\pi}\right)^2
	=\frac{1}{8\pi^2\Mp^2}\frac{\dot{\phi}^2}{\epsilon_1 c_s}.
\eea
Therefore, with use of the relations $\partial_\phi = \partial_t/\dot{\phi}$ and $N_\phi = -H/\dot{\phi}$, one obtains
\bea
	\partial_\phi\log\mathcal{P}_{\delta\phi}=N_\phi\left(\epsilon_2-2\delta+s\right),
\eea
and \Eq{k-inflation} is recovered. In fact, the result~(\ref{k-inflation:fnltotal}) should not come as a surprise since when dealing with the single-field case in \Eqs{eq:forward:SF:deltaNlPzetaS}--(\ref{eq:forward:SF:fnl}), no assumption was made regarding $\calP_{\delta\phi}$ and the CR was therefore also valid in the case of  non-canonical kinetic terms.
\subsubsection{Backward formulation}
\label{sec:backward formulation}
\begin{figure}[t]
\begin{center}
\includegraphics[width=0.7\textwidth]{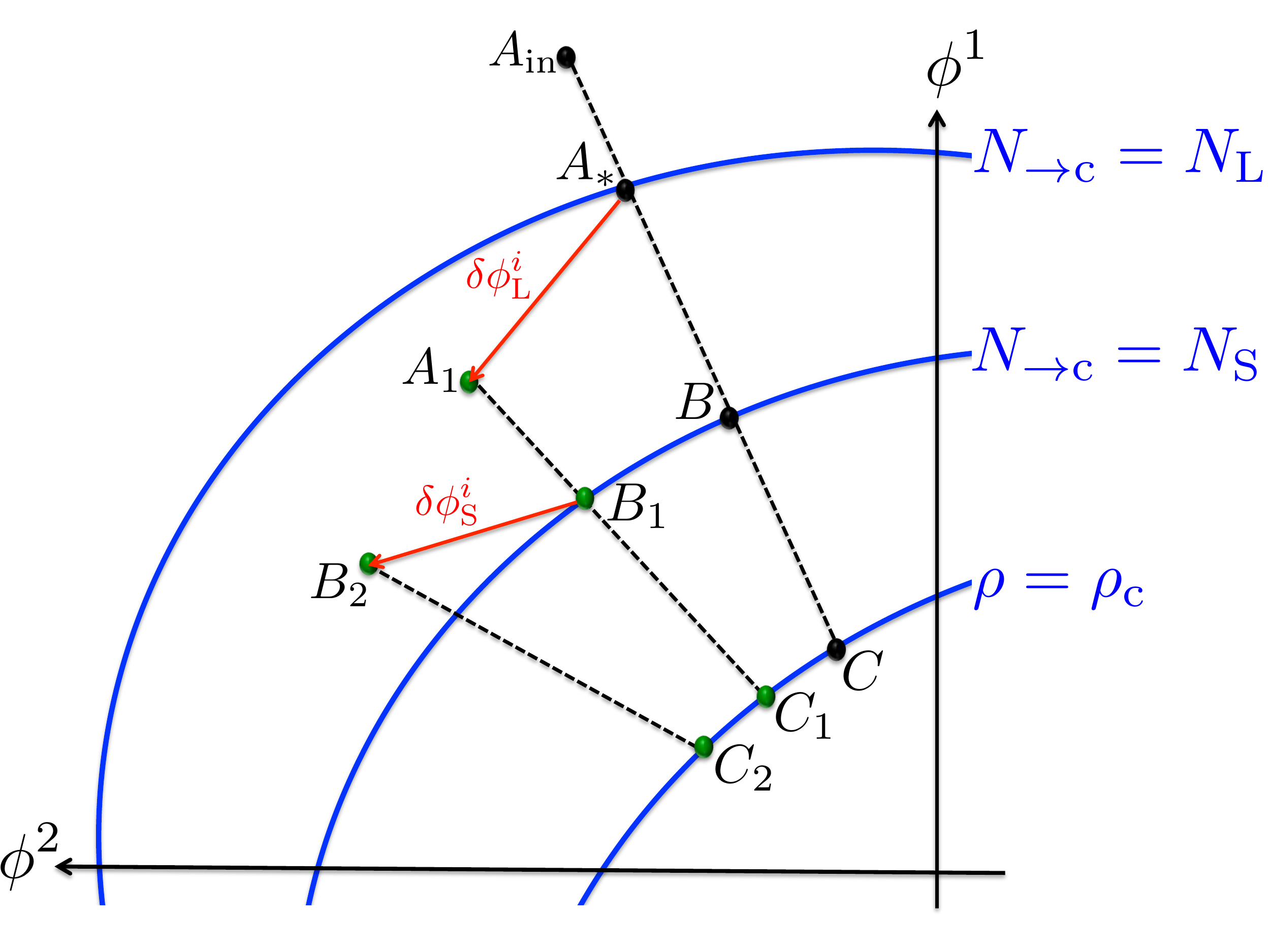}
\caption{Schematic representation of the backward procedure for the calculation of the squeezed bispectrum in the $\delta N$ formalism. In this formulation, the backward number of \efolds $N_{\to \mathrm{c}}$ determines the location of $B_1$, $N_{B_1 C_1} = N_\uS$. This condition yields unperturbed physical scales on the uniform density hypersurface $\rho=\rho_\uc$. }
\label{fig:schema}
\end{center}
\end{figure}
The forward formulation developed in \Sec{sec:forwardformulation} gives the standard bispectrum $B_\zeta$ in terms of comoving scales. Let us now derive the bispectrum $B_{\bar{\zeta}}$ in terms of the physical scales that would be seen by a local observer. 
The idea is to define all perturbation scales using the \emph{backward} number of \efolds realised until the final uniform density hypersurface is reached. 
This procedure is thus called ``backward formulation'' and is summarised in \Fig{fig:schema}. 
Since the physical Hubble scale $H$ on the final uniform density slice is constant, the Hubble crossing physical scale $k_\uc^\mathrm{phys}$ is constant on this hypersurface as well, which implies that the physical scales $k_\uL^\mathrm{phys} = \ee^{-N_\uL}k_\uc^\mathrm{phys}$ and $k_\uS^\mathrm{phys} = \ee^{-N_\uS}k_\uc^\mathrm{phys}$ are also unperturbed quantities on the final uniform density slice.

From \Figs{fig:schemaclass} and~\ref{fig:schema}, one can see that the only difference between the forward and backward formulations is the definition of the point $B$.
Therefore, the formal expression of $f_\mathrm{NL}$ given in \Eq{fNL beyond equilateral} still applies here. 
Before explaining how it can be evaluated in general, let us see how it compares with the forward formulation in the single-field case and in the near-equilateral limit respectively.\\

\noindent\textbf{Single-field case}\\
In single-field inflation, one can readily see that the backward formulation always yields a vanishing squeezed bispectrum. In this case indeed, the hypersurfaces of constant backward number of \efolds (the blue lines in \Fig{fig:schema}) are single points, and the point $B_1$ always coincides with $B_*$ irrespectively of the long-wavelength perturbation $\delta\phi_\uL $. There is therefore no correlation between the long-wavelength perturbation at $A_*$ and the power spectrum at $B_1$, and the squeezed bispectrum vanishes. In this picture, the fact that the squeezed non-linearity parameter vanishes in single-field inflation has a clear geometrical interpretation in field space.\\

\noindent\textbf{Near-equilateral limit}\\
\begin{figure}[t]
\begin{center}
\includegraphics[width=0.3\textwidth]{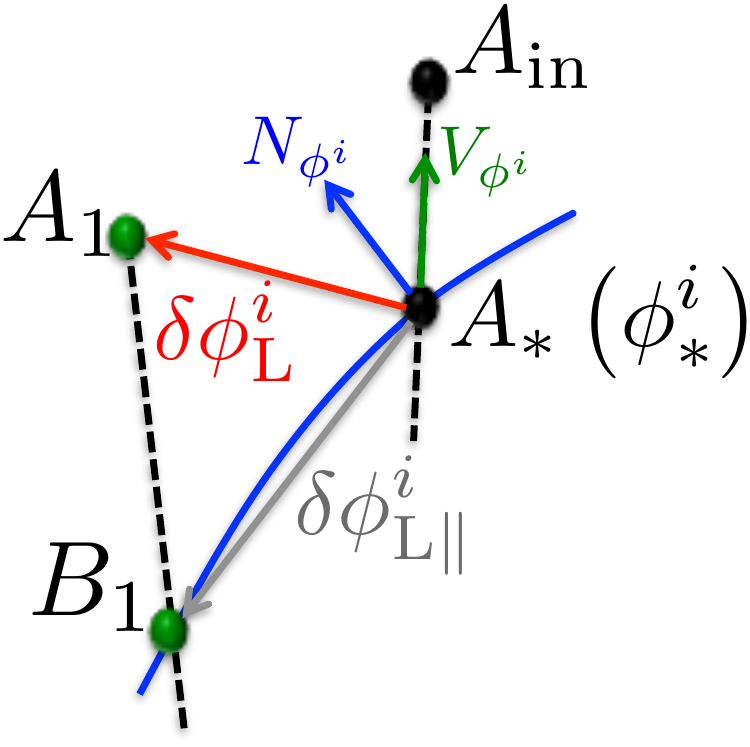}
\caption{Schematic representation of the backward formulation for the calculation of the squeezed bispectrum in the near-equilateral limit. In this regime, the constant backward number of \efolds hypersurfaces $N_{\to \mathrm{c}}=N_\uL $ and $N_{\to \mathrm{c}}=N_\uS $ coincide (blue line). The slow-roll trajectories $[A_\mathrm{in}A_*]$ and $[A_1B_1]$ are aligned with the gradient of the potential $V_i$. The gradients of the backward number of \efolds function $N(\phi)$ and of the potential $V(\phi)$ are misaligned in general, but if they are parallel (as in single-field inflation), the squeezed bispectrum vanishes.}
\label{fig:schemaNlNs}
\end{center}
\end{figure}
Let us now work out the near-equilateral limit $N_\uS \to N_\uL $, illustrated in \Fig{fig:schemaNlNs}. It corresponds to the limit $A_*\to B$ but note that the point $A_1$ does not coincide with $B_1$ in contrast to the forward formulation.
The difference in the field values between $A_*$ and $B_1$ is denoted $\delta\phi_{\uL||}^i$. In the limit $N_\uS \to N_\uL $, the numbers of \efolds realised until the uniform density hypersurface is reached are the same from $A_*$ and from $B_1$, so that
\bea
\label{eq:backward:equilateral:C1}
	0=N\left({\phi}_*^j+\delta{\phi}_{\uL ||}^j\right)-N({\phi}_*^j)\simeq N_i\left({\phi}_*\right)\delta\phi_{\uL ||}^i.
\eea
Another condition to write down is that the points $A_1$ and $B_1$ lie on the same slow-roll trajectory, which implies that $[A_1 B_1]$ is parallel to the gradient of the potential, giving rise to
\bea
\label{eq:backward:equilateral:C2}
	\delta\phi_\uL ^i-\delta\phi_{\uL ||}^i=\lambda V_i,
\eea
where $\lambda$ is a constant. Combining \Eqs{eq:backward:equilateral:C1} and~(\ref{eq:backward:equilateral:C2}), one can solve both for $\lambda={N_i\delta\phi_\uL ^i}/({N_jV_j})$ and for 
\bea
	\delta\phi_{\uL ||}^i=\delta\phi_\uL ^i-\frac{N_j\delta\phi_\uL ^j}{N_kV_k}V_i.
\eea
The power spectrum at $B_1$ can then be evaluated as
\bea
	\left.\mathcal{P}_\zeta\right\vert_{B_1}&\simeq & \mathcal{P}_\zeta({\phi}_*)+\pdif{\mathcal{P}_\zeta}{\phi^j}\delta\phi_{\uL ||}^j \nonumber \\
	&\simeq & \mathcal{P}_\zeta({\phi}_*)+2\left(\frac{H}{2\pi}\right)^2\left(N_iN_{ij}+N_iN_i\frac{H_j}{H}\right)
	\left(\delta\phi_\uL ^j-\frac{N_k\delta\phi_\uL ^k}{N_lV_l}V_j\right),
\eea
and its correlation with $\delta N_\uL \simeq N_i\delta\phi_\uL ^i$ is given by
\bea
	\Braket{\delta N_\uL \mathcal{P}_\zeta(k_\uS )}&=&2\left(\dfrac{H}{2\pi}\right)^2N_i\left(N_jN_{jk}+N_jN_j\dfrac{V_k}{2V}\right)
	\left(\Braket{\delta\phi_\uL ^i\delta\phi_\uL ^k}
	-\dfrac{N_l\Braket{\delta\phi_\uL ^i\delta\phi_\uL ^l}}{N_mV_m}V_k\right) \nonumber \\
	&=&2\left(\dfrac{H}{2\pi}\right)^4N_i\left(N_jN_{jk}+N_jN_j\dfrac{V_k}{2V}\right)\left(\delta_{ik}
	-\frac{N_l\delta_{il}}{N_mV_m}V_k\right).
\eea
Plugging this expression into \Eq{fNL w/ impulsive fluctuation}, one obtains for the squeezed non-linearity parameter
\bea
	\frac{3}{5}\fnl=\frac{1}{2(N_mN_m)^2}\left(N_jN_{ij}+N_jN_j\dfrac{V_i}{2V}\right)\left(N_i-\frac{N_lN_l}{N_kV_k}V_i\right).
	\label{eq:fNL:backwards:std:SeqL}
\eea
This formula is one of the main results of this paper since it allows one to directly calculate the squeezed $\fnl$ parameter (in the near-equilateral limit) in terms of physical scales and as seen by a local observer.

Let us note that the expression~(\ref{eq:fNL:backwards:std:SeqL}) derived in the backward formalism is similar to the one~(\ref{fNL std}) derived the forward formalism, except that $N_i$ is now replaced by $N_i-\frac{N_lN_l}{N_kV_k}V_i$. Therefore, $\fnl$ in the squeezed formalism is proportional to the projection of the gradient of $V$ on the hypersurface of constant backward number of $e$-folds. In other words, if the gradients of $V(\phi)$ and $N(\phi)$ in field space are parallel, then the squeezed bispectrum vanishes. This is obviously the case in single-field inflation as already mentioned, but can also happen in some multiple-field models. 
Let us also mention that\footnote{The scalar spectral
index is given by
\bea\label{1-ns}
	1-\nS=\dif{\log\mathcal{P}_\zeta}{N}=\frac{1}{\mathcal{P}_\zeta}\dif{\phi^i}{N}\partial_i\mathcal{P}_\zeta
	=\frac{\Mp^2}{N_kN_k}\frac{V_i}{V}\left(2N_jN_{ij}+N_jN_j\frac{V_i}{V}\right)\, ,
\eea
where the slow-roll equation of motion~(\ref{eq:SRtraj}) has been used.
Let us consider two near constant backward number of \efolds hypersurfaces $N$ and $N+\delta N$ and let $\delta\bar{\phi}^i$ denote the field variation between these two slices along the unperturbed attractor trajectory.
By definition of $N_i$, one has $\delta N=N_i\delta\bar{\phi}^i$. On the other hand, the slow-roll equation of motion~(\ref{eq:SRtraj}) for $\phi^i$ yields $\delta\bar{\phi}^i=\Mp^2\frac{V_i}{V}\delta N$. Combining these two relations, one obtains
\bea
	\Mp^2N_iV_i=V,
\eea
which allows one to show that the difference between the result of the forward formulation~(\ref{fNL std}) and the one of the backward formulation~(\ref{eq:fNL:backwards:std:SeqL})
is indeed given by $\frac{1-\nS}{4}$.}  the difference between the forward and the backward formulas is simply given by $\frac{3}{5}\fnl^\mathrm{CR}=\frac{1-\nS}{4}$, in agreement with \Eq{eq:fnc:squeezed3pt:trans}. In \Sec{sec:examples}, this consistency check will be extended beyond the near-equilateral limit.
\subsection{Computational programme}
\label{sec:ComputationalProgramme}
Let us now explain how the approach proposed in the present work can be implemented in practice. The formula~(\ref{fNL w/ impulsive fluctuation}) allows one to calculate the squeezed $\fnl(\kL,\kS)$ parameter 
both in the forward and in the backward formulation for any multi-field model of inflation. 
In general 
the quantities appearing in \Eq{fNL beyond equilateral} cannot be calculated analytically and one has to evaluate them numerically, according to the following procedure:
\begin{enumerate}
\item \label{CP:1} Starting from $A_\uin(\phi^i_\uin)$, integrate the background equation of motion for the fields and find the coordinates in field space of $A_*$, $B$ and $C$ defined in \Fig{fig:schemaclass} (forward) or \ref{fig:schema} (backward).
\item \label{CP:2} Starting from $\phi_*^j+\epsilon\delta_i^j$, integrate the background equation of motion until the condition $\rho=\rho_\uc$ is met. If $N^i_\epsilon$ denotes the realised number of $e$-folds, assess $N_{i}\vert_*=(N^i_\epsilon-N_\uL )/\epsilon$. Repeat this step for $i=1\cdots D$, where $D$ is the number of fields.
\item \label{CP:3} Reproduce step~\ref{CP:2} but starting from $B$ instead of $A_*$ to compute $N_{i}\vert_B$.
\item \label{CP:4} Compute the power spectrum amplitude $\calP_\zeta=N_iN_j\langle\delta\phi^i_{\uL\,(\uS)}\delta\phi^j_{\uL\,(\uS)}\rangle$
at $A_*$ and at $B$. In this expression, the derivatives $N_i$ have been computed in step~\ref{CP:2} for $A_*$ and in step~\ref{CP:3} for $B$, but $\smash[b]{\langle\delta\phi^i_{\uL\,(\uS)}\delta\phi^j_{\uL\,(\uS)}\rangle}$
depends on the model one considers. In standard slow-roll inflation for instance, it is simply given by $\smash[b]{\langle\delta\phi^i_{\uL\,(\uS)}\delta\phi^j_{\uL\,(\uS)}\rangle =\calP_{\delta\phi}\delta_{ij}= [H/(2\pi)]^2\delta_{ij}}$
and is straightforward to compute.
\item \label{CP:5} Using the trajectories integrated in step~\ref{CP:2}, find the coordinates of $B_\epsilon^i$ defined with respect to the shifted starting points $\phi_*^j+\epsilon\delta_i^j$. Using the same technique as in step~\ref{CP:4}, compute $\calP_{\zeta} (B_\epsilon^i)$, the power spectrum amplitude at $B_\epsilon^i$. Assess $\partial\calP_{\zeta}(B)/\partial\phi_*^i = [\calP_{\zeta} (B_\epsilon^i)-\calP_{\zeta} (B)]/\epsilon $, where $\calP_{\zeta} (B)$ has been computed in step~\ref{CP:4}. Repeat this step for $i=1\cdots D$.
\item \label{CP:5} Evaluate
\bea
\frac{3}{5}\fnl = \dfrac{\left.N_{i}\right\vert_*\dfrac{\partial\calP_{\zeta} (B)}{\partial\phi^i_*}\left.\calP_{\delta\phi}\right\vert_*}{4\calP_\zeta(A_*)\calP_{\zeta}(B)},
\eea
where $N_{\phi^i_*}$ has been computed in step~\ref{CP:2}, $\partial\calP_{\zeta}( B)/\partial\phi^i_*$ in step~\ref{CP:5}, and $\calP_\zeta (A_*)$ and $\calP_{\zeta} (B)$ in step~\ref{CP:4}. 
\end{enumerate}
This makes the implementation of our proposal straightforward as soon as one knows how to solve the background equation of motion. In the next section, we discuss the results it gives in two examples.
\section{Examples}
\label{sec:examples}
In the previous section, we have explained how to calculate the squeezed $\fnl(\kL,\kS)$ parameter in a generic multiple-field model of inflation, in the forward and in the backward formulation, allowing for an arbitrary separation between the two scales $\kL$ and $\kS$, and automatically taking the intrinsic NG component into account. Let us now illustrate our approach on two concrete two-field models. 
In the first one, the two fields have a simple quadratic potential and in the second one, inflation is driven by a single field but its value at the end of inflation explicitly depends on a second field. This shows how the method we propose works in practice, and provides a few interesting results for the value taken by the squeezed non-linearity parameter in these setups.
We focus on the backward formulation since it yields the result a local observer would see, and given that, as already pointed out, both formulations only differ by $\frac{1-\nS(k_\uS)}{4}$.
\subsection{Double massive inflation}
\label{sec:doublemassive}
Let us consider the case where inflation is driven by two scalar fields $\phi$ and $\psi$, slowly rolling down the potential
\bea
\label{eq:doublemassive:pot}
V(\phi,\psi)=\dfrac{m_\phi^2}{2}\phi^2 + \dfrac{m_\psi^2}{2}\psi^2.
\eea
In the following, without loss of generality, we assume that $m_\phi\geq m_\psi$. In this case inflation is first mainly driven by $\phi$, before a turning point occurs in field space and $\psi$ takes over. The ending hypersurface is determined by $\rho=\rho_\uc$, where $\rho_\uc$ is the energy density when $\epsilon_1=1$ on the unperturbed trajectory.

In \Sec{sec:ComputationalProgramme}, we have sketched the computational program that allows one to numerically evaluate $\fnl$. For the model~(\ref{eq:doublemassive:pot}), there exists an analytical integral of motion $K=m_\psi^2\ln\phi - m_\phi^2\ln\psi$ that is constant along slow-roll trajectories and that can label them~\cite{Sasaki:2008uc}. Moreover, the number of \efolds realised between two points $M_1(\phi_1,\psi_1)$ and $M_2(\phi_2,\psi_2)$ in field space that belong to the same slow-roll trajectory (\ie that share the same value of $K$) is given by $N_{12}=(\phi_1^2+\psi_1^2-\phi_2^2-\psi_2^2)/(4\Mp^2)$~\cite{Starobinsky:1986fxa}. These relations provide analytical results for all steps of the procedure described in \Sec{sec:ComputationalProgramme}. This is in fact the case for all separable potentials (either additive separable as in here or multiplicative separable), for which we provide all relevant formulas in \Apps{sec:sepAddPot} and~\ref{sec:sepMulPot}. 
\begin{figure}[t]
	\begin{center}
	\includegraphics[width=\wdblefig]{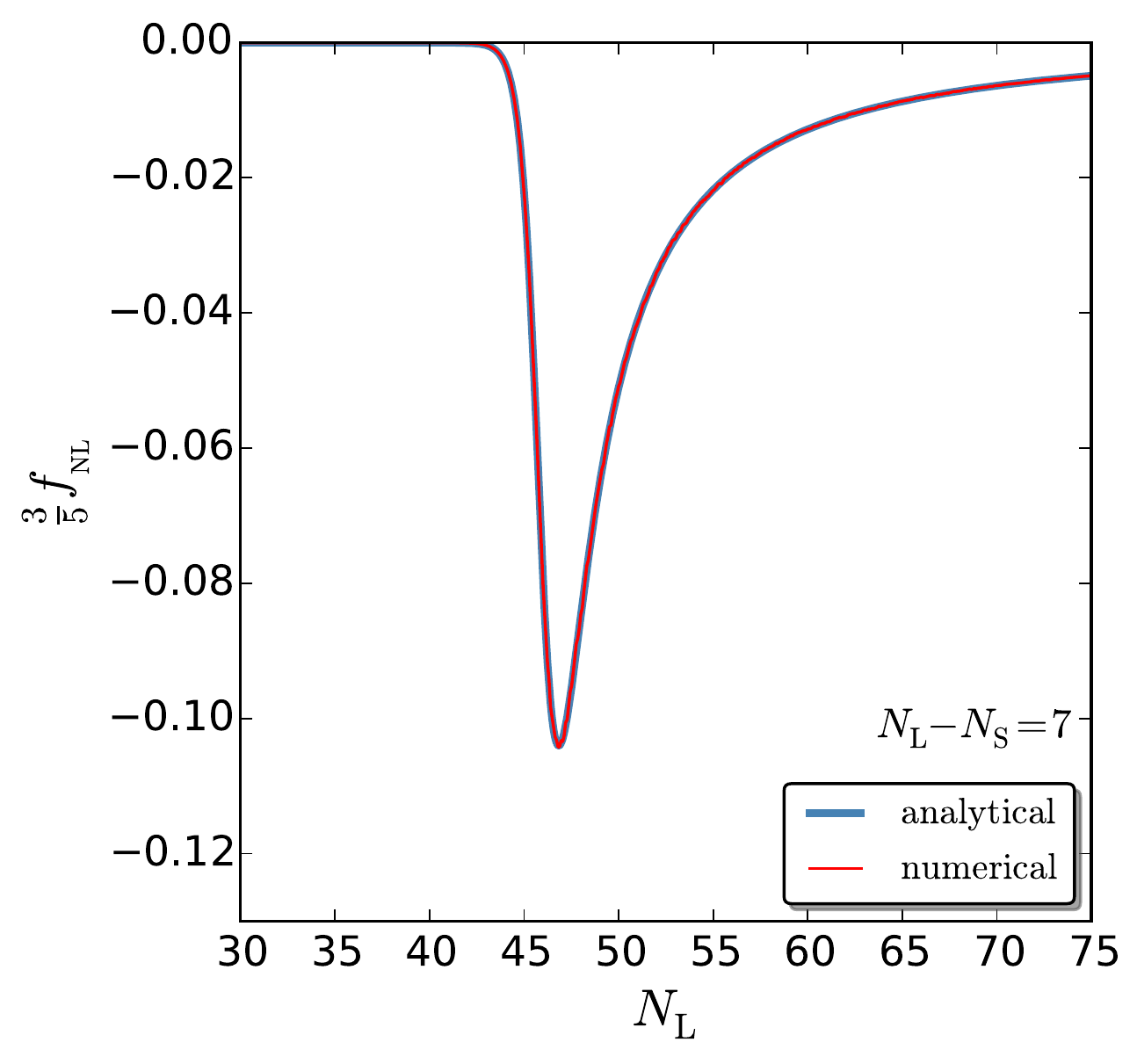}
	\includegraphics[width=1.0\wdblefig]{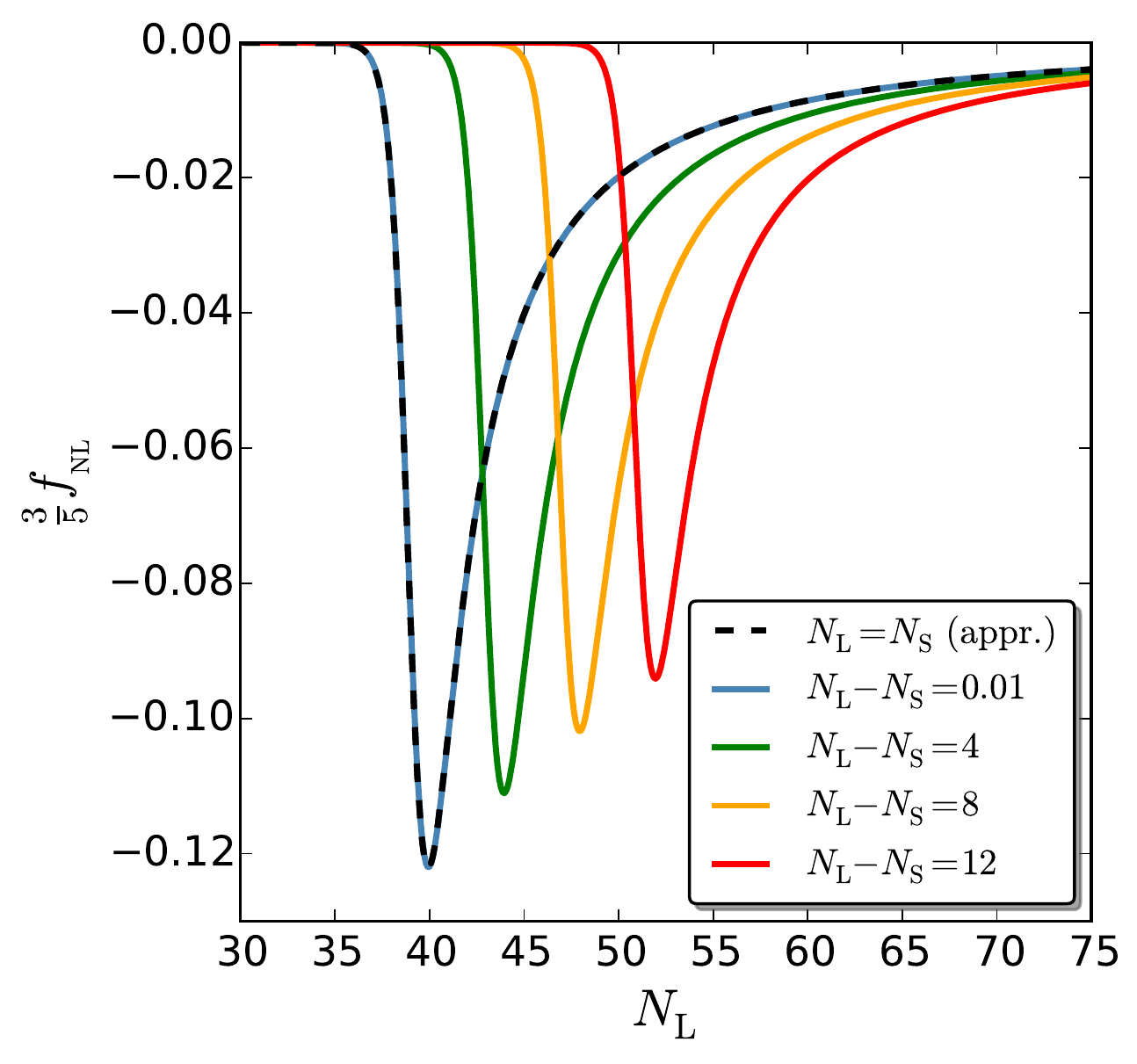}
	\caption{Squeezed backward $\fnl$ parameter, computed in the double massive potential~(\ref{eq:doublemassive:pot}), as a function of $N_\uL $, with $\phi_\uin=\psi_\uin=13\Mp$, $m_\phi/\Mp=9\times 10^{-6}$, $m_\psi=m_\phi/9$ and ${\rho}_\uc=m_\psi^2\Mp^2$ (corresponding to the value of $\rho$ at the end of slow-roll inflation on the background trajectory associated with $\phi_\uin$ and $\psi_\uin$), matching the parameters used in the figures of \Refs{Rigopoulos:2005ae, Vernizzi:2006ve}. In the left panel, $N_\uS =N_\uL -7$, and a comparison between the analytical formulas of \App{sec:sepAddPot} (blue) and the numerical procedure described in \Sec{sec:ComputationalProgramme} (red) is displayed. One can check that the agreement is excellent. In the right panel, the result is shown for a few values of $N_\uL -N_\uS $ (labeled by different colours). The black dashed line stands for the limit $N_\uS  \rightarrow N_\uL $ given by \Eq{eq:doublemassive:SeqL} where the contributions from the end of inflation are neglected. One can check that when $ N_\uL  - N_\uS  \ll 1$, this provides a good approximation indeed.}
	\label{fig:DoubleMassive}
	\end{center}
\end{figure}
In the left panel of \Fig{fig:DoubleMassive}, we compare the result of the backward formulation 
obtained from these analytical formulas with the numerical procedure described in \Sec{sec:ComputationalProgramme}. One can check that the agreement is excellent, confirming both approaches.

In the right panel of \Fig{fig:DoubleMassive}, different values of $N_\uL -N_\uS $ are displayed and a few remarks are in order. First, one can see that  $\vert \fnl \vert$ reaches a maximum, corresponding to when the scales one considers exit the Hubble radius at the time when the turn in field space is maximal (which happens about $40$ \efolds before the end inflation for the parameters used in \Fig{fig:DoubleMassive}). Away from this point, the model effectively describes a single-field system (driven by $\phi$ much before the turning point and by $\psi$ much after the turning point) and $\fnl$ vanishes. Second, the value of $N_\uL $ at which $\vert \fnl \vert$ is maximal is shifted by $N_\uL -N_\uS $ from one curve to the other. This indicates that, in fact, $\vert \fnl \vert$ is maximal if $k_\uS $, the smaller wavelength, exits the Hubble radius at the time of maximal turn in field space. Third, when $N_\uL -N_\uS $ increases, the maximal value of $\vert \fnl \vert$ decreases. This can be understood noticing that fixing $N_\uS $ to the time of maximal turn in field space, the large wavelength fluctuation exits the Hubble radius away from the turning point if $N_\uL -N_\uS $ is large, that is to say when the system is effectively single field. At this stage, the gradients of $N$ and $V$ (in field space) are nearly aligned, and the fluctuation $\delta N_\uL $ almost does not change the slow-roll trajectory along which the system evolves. Hence it almost does not correlate with $\calP_\zeta(\kS)$, leading to a small $\fnl$. More generally, it confirms that it is important to account for the actual values of the scales $k_\uL$ and $k_\uS$ to properly compute $\fnl$, and that the near-equilateral limit $k_\uL \to k_\uS$ does not always provide a reliable estimate when the two scales differ.

In this model, $\fnl$ is therefore maximal when the two scales $\kL$ and $\kS$ are close. In this limit, if one ignores the contribution from the end of inflation and simply plugs $N=(\phi^2+\psi^2)/(4\Mp^2)+\mathrm{constant}$ into \Eq{eq:fNL:backwards:std:SeqL}, one obtains
\bea
\frac{3}{5}\fnl(\kS\rightarrow\kL) \simeq \frac{\Mp^2}{\phi^2+\psi^2}\left[1-\frac{\left(\phi^2+\psi^2\right)\left(m_\phi^4\phi^2+m_\psi^4\psi^2\right)}{\left(m_\phi^2\phi^2+m_\psi^2\psi^2\right)^2}\right].
\label{eq:doublemassive:SeqL}
\eea
In this expression, one can check that $m_\phi=m_\psi$ leads to $\fnl=0$, which is consistent with the fact that when the two masses are equal, the model is effectively equivalent to a single-field setup and $\fnl$ vanishes. This expression is displayed in the right panel of \Fig{fig:DoubleMassive} as the black dashed line. One can check that when $ N_\uL  - N_\uS  \ll 1$, it provides a good approximation to the exact result. The reason is that if the turn in field space occurs much before the end of inflation, inflation is effectively driven by one of the fields only when it ends and the vev of the other field almost vanishes. In this limit, it is safe to neglect the dependence of the end of inflation field space coordinates  on small changes in the initial conditions.
\begin{figure}[t]
	\centering
	\begin{tabular}{cc}
		\begin{minipage}{0.491\hsize}
			\centering
			\includegraphics[width=\hsize]{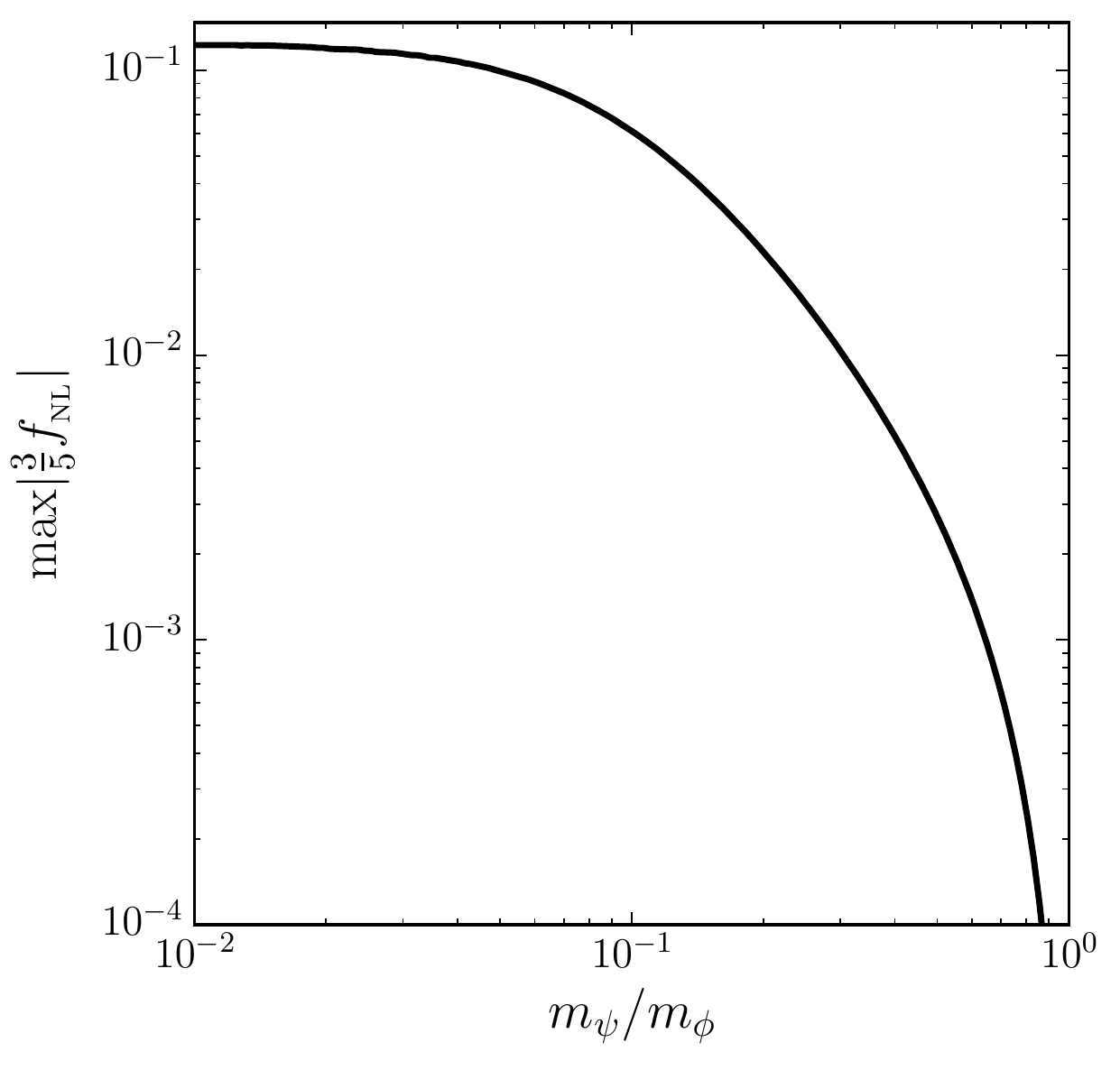}
		\end{minipage}
		\begin{minipage}{0.509\hsize}
			\centering
			\vspace{-13pt}
			\includegraphics[width=\hsize]{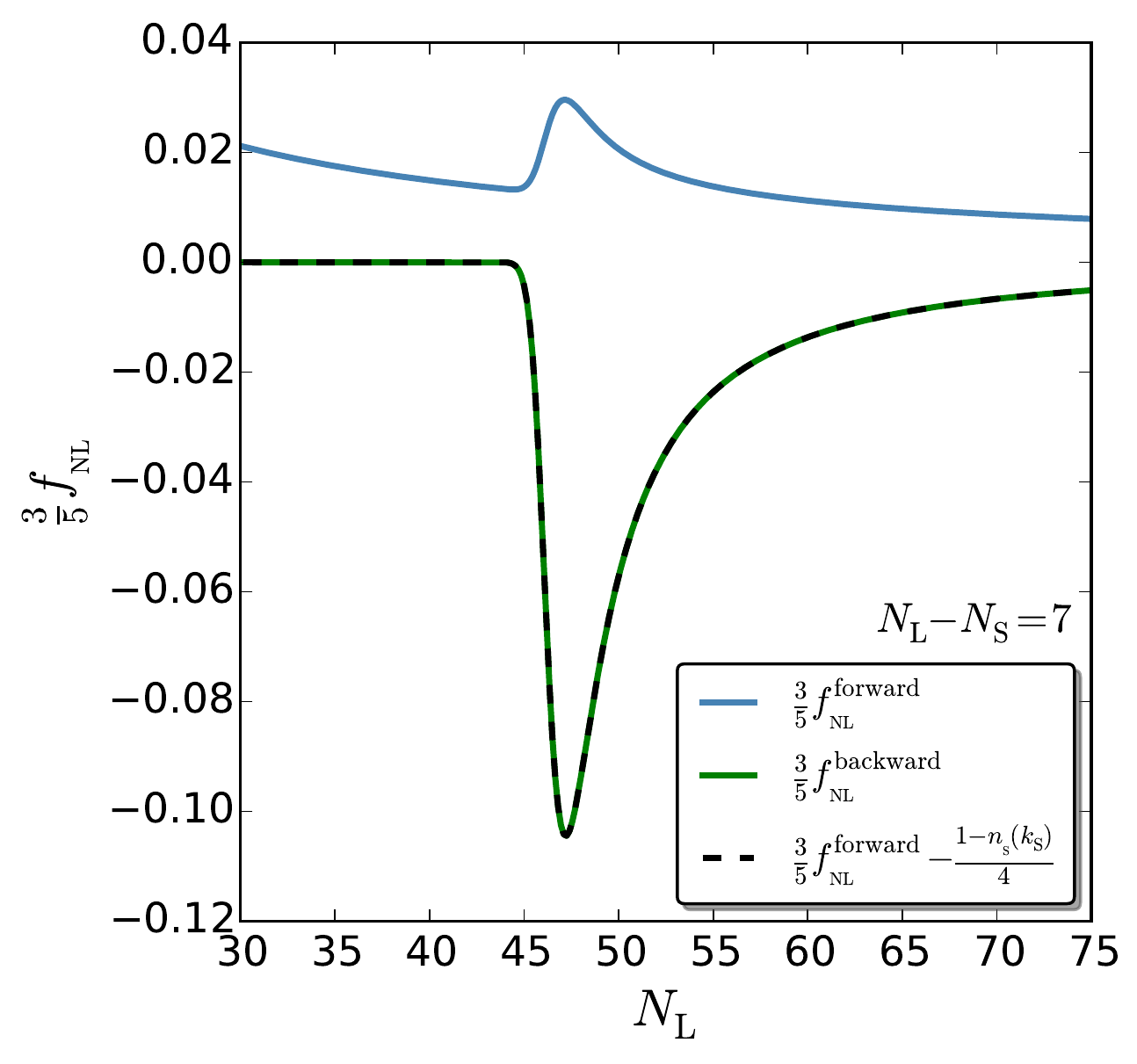}
		\end{minipage}
	\end{tabular}
	\caption{Left panel: maximum squeezed $\fnl$ parameter, computed in the double massive potential~(\ref{eq:doublemassive:pot}), 
	as a function of $m_\psi/m_\phi$, with $N_\uL =N_\uS =50$, where maximisation is performed over all trajectories in field space.
	Right panel: Forward and backward formulations results for squeezed $\fnl$ as a function of $N_\uL$, where $N_\uL-N_\uS$ is fixed to $7$. The black dotted line shows 
	$\frac{3}{5}f_\mathrm{NL}^\mathrm{forward}-\frac{1-\nS(k_\uS)}{4}$ where the spectral index is given by \Eq{1-ns}.
	It matches $\frac{3}{5}f_\mathrm{NL}^\mathrm{backward}$ and confirms that even 
	beyond the near-equilateral limit, the backward formulation yields the squeezed bispectrum in terms of physical scales as seen by a local observer.}
	\label{fig:DoubleMassive:fnlmax}
\end{figure}

In the limit $\kS\rightarrow\kL$, \Eq{eq:doublemassive:SeqL} also indicates that $\fnl$ only depends on the ratio between the two masses $m_\phi$ and $m_\psi$, and not on the absolute values of the masses. For this reason, in the left panel of \Fig{fig:DoubleMassive:fnlmax}, 
the maximal value of $\fnl$ is displayed as a function of $m_\psi/m_\phi$, where the maximisation is performed over all possible trajectories in field space and $\fnl$ is computed $50$ \efolds before the end of inflation. One can see that in this model, the squeezed $\fnl$ parameter can never be larger than of order $\order{0.1}$.

Finally, let us see how this result compares to the forward formulation. In the right panel of \Fig{fig:DoubleMassive:fnlmax},
the non-linearity parameter $f_\mathrm{NL}$ is displayed in each formulation for $N_\uL-N_\uS=7$. It is interesting to notice that opposite signs are obtained with the two formulations. Moreover, one can check that $\frac{3}{5}\fnl^\mathrm{forward}-\frac{1-\nS(k_\uS)}{4}$, represented by the black dotted line, matches $\frac{3}{5}\fnl^\mathrm{backward}$. This is in agreement with \Eq{eq:fnc:squeezed3pt:trans} and confirms that the backward formulation yields the squeezed bispectrum in terms of physical scales as seen by a local observer.

\subsection{Inhomogeneous end of inflation}
Let us further illustrate how the calculational program proposed in this work can be implemented with the example of inhomogeneous end of inflation models. In such setups, inflation is driven by a single field $\phi$ with potential $V(\phi)$, but inflation ends at a value $\phi_C(\psi)$ that depends on the vev of a second field $\psi$. The calculation of $\fnl$ for this class of models is presented in Appendix~\ref{sec:inhom}. In this section, we consider the case where $V(\phi)$ is exponential and $\phi_C(\psi)$ is a circular function
\bea
\label{eq:inhom:model}
	V(\phi) = V_0 \exp\left(\alpha\frac{\phi}{\Mp}\right), \quad \phi_C(\psi) = \mu\cos\left(\frac{\psi}{\psi}_0\right).
\eea
In this model, the number of $e$-folds~(\ref{eq:NM1M2:inhom}) can easily be worked out, $N_{M_1M_2}=(\phi_1-\phi_2)/(\alpha\Mp)$, 
and the coordinates of $B$ in the backward formulation are given by $\phi_B = \mu\cos(\psi/\psi_0) + \alpha\Mp N_\uS $ and $ \psi_B=\psi$.

With these formulas, the procedure detailed in Appendix~\ref{sec:inhom} gives rise to
\bea
\frac{3}{5}\fnl^\mathrm{backward}&=\dfrac{\alpha}{4}\dfrac{\mu^2}{\psi_0^2}\sin^2\left(\dfrac{\psi}{\psi_0}\right)
\dfrac{\left[2\dfrac{\Mp\mu}{\psi_0^2}\cos\left(\dfrac{\psi}{\psi_0}\right)-
\alpha-\alpha\dfrac{\mu^2}{\psi_0^2}\sin^2\left(\dfrac{\psi}{\psi_0}\right)
\right]}
{\left[1+\dfrac{\mu^2}{\psi_0^2}\sin^2\left(\dfrac{\psi}{\psi_0}\right)\right]^2}.
\eea
In this expression, let us note that $\fnl$ neither depends on $N_\uL $ nor $N_\uS $, but only on $\psi$, which is constant during inflation. Contrary to the example discussed in \Sec{sec:doublemassive}, the limit $k_\uS \rightarrow k_\uL $ is therefore not restrictive since the result does not depend on $k_\uS $ and $k_\uL $. 

\begin{figure}[t]
	\begin{center}
	\includegraphics[width=0.46\hsize]{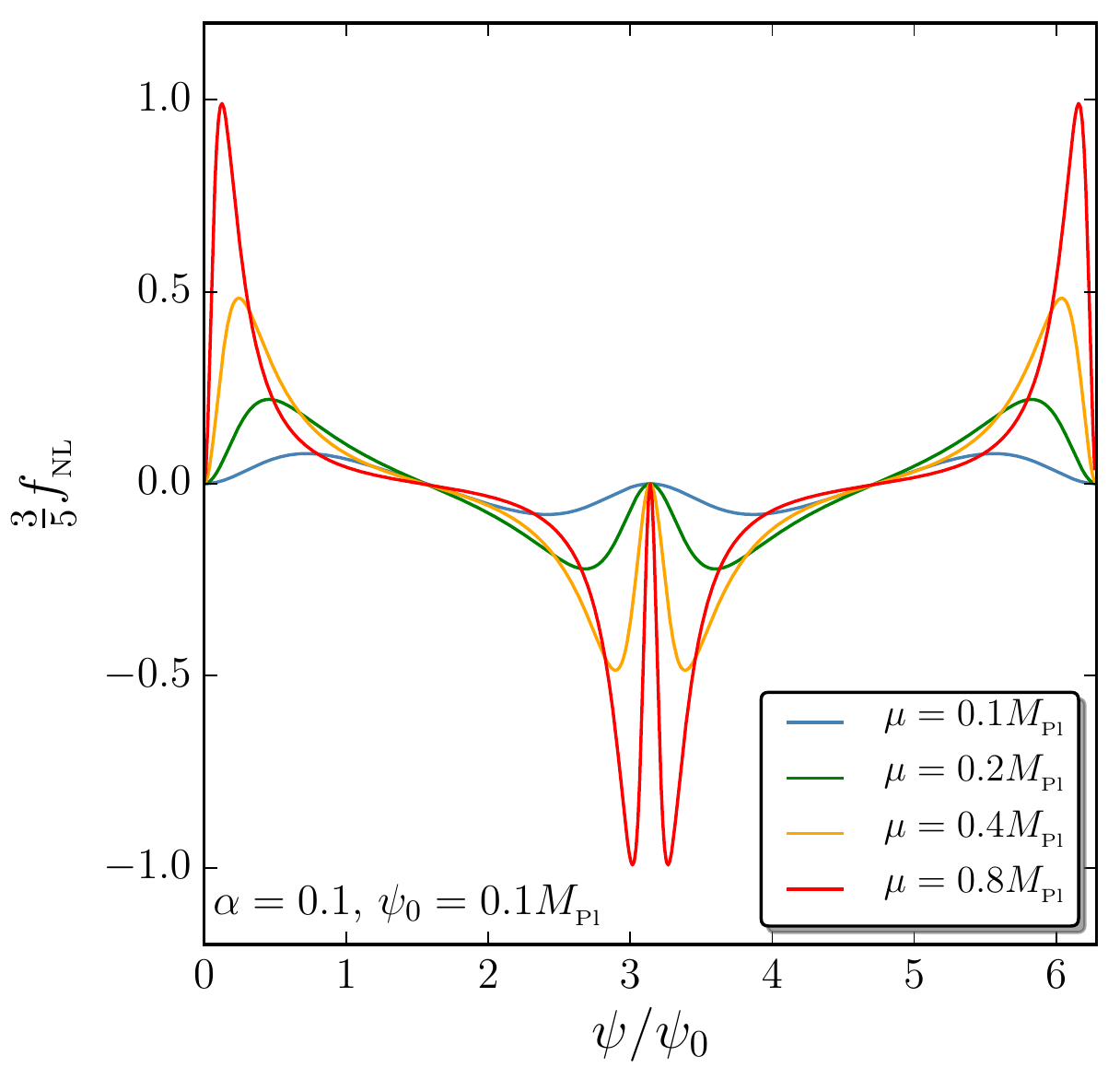}
	\includegraphics[width=0.505\hsize]{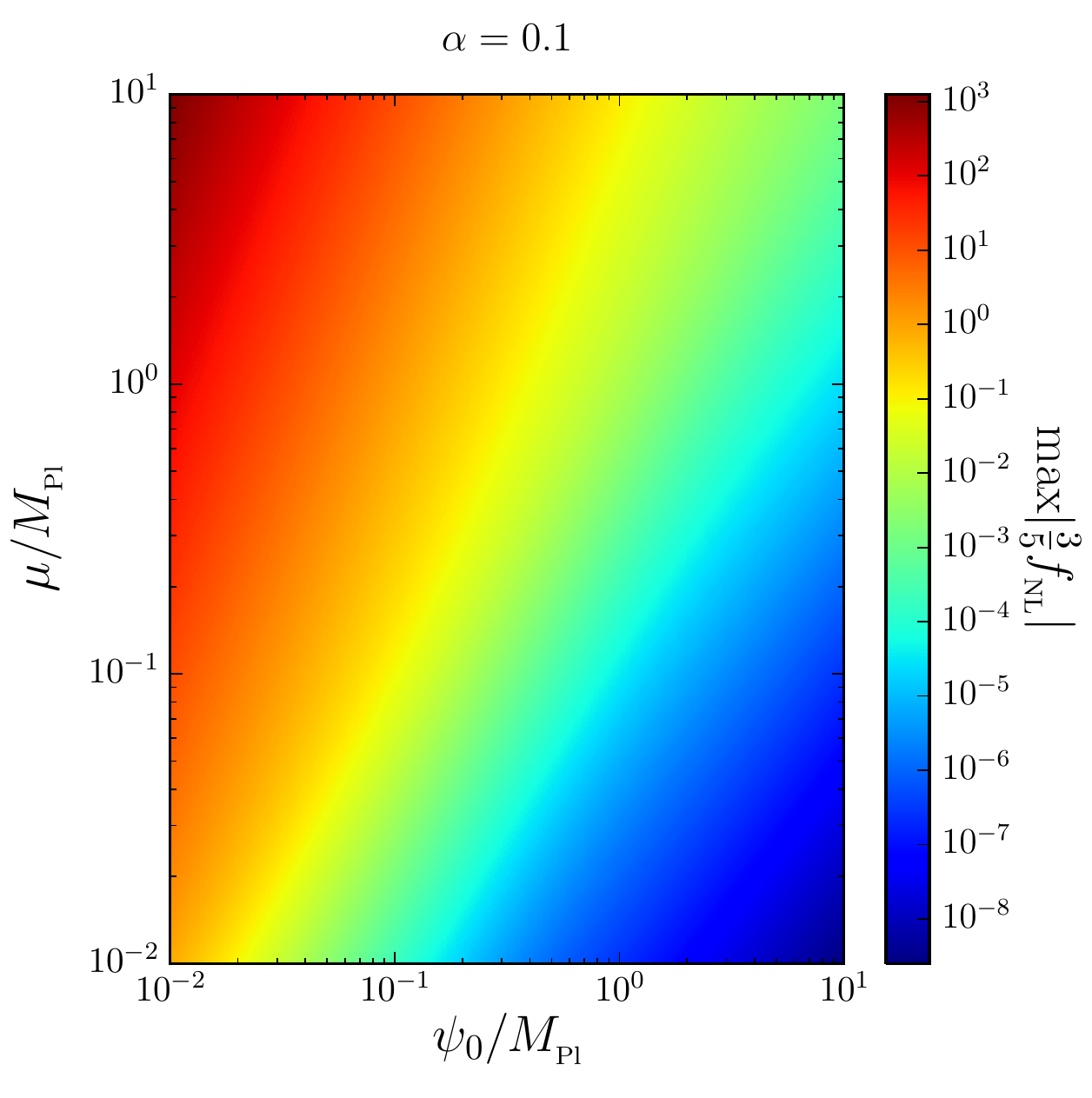}
	\caption{Squeezed backward $\fnl$ parameter in the inhomogeneous end of inflation model~(\ref{eq:inhom:model}). In the left panel, $\fnl$ is plotted as a function of $\psi/\psi_0$ with $\alpha=\psi_0/\Mp=0.1$, and a few values of $\mu$ are displayed. In the right panel, the maximal value of $\fnl$ over $\psi$ is shown, as a function of $\psi_0$ and $\mu$, with $\alpha=0.1$. Large values of $\vert\fnl\vert$ can be obtained when $\mu$ is large and $\psi_0$ is small.}
	\label{fig:inhom}
	\end{center}
\end{figure}
In the left panel of \Fig{fig:inhom}, the backward $\fnl$ parameter is displayed as a function of $\psi$ for $\alpha=0.1$, $\psi_0/\Mp=0.1$ and a few values of $\mu$.  When $\psi/\psi_0=n\pi$, where $n$ is an integer number, $\fnl$ vanishes since $\partial N/\partial\psi = 0$. Otherwise it can take positive or negative values depending on $\psi$. In the right panel of \Fig{fig:inhom}, the maximal value of $\vert \fnl\vert$ over $\psi$ is shown as a function of $\psi_0$ and $\mu$, for $\alpha=0.1$. One can see that a large squeezed $\fnl$ parameter can be obtained if $\psi_0$ is small and $\mu$ is large, since in this case the end surface of inflation is oscillating with a large amplitude and a large frequency. Let us finally mention that these results are valid in the regime where the quantum spread of $\psi$ at the end of inflation, $\sim\sqrt{50}H/(2\pi)$, is much smaller than the period of $\phi_C(\psi)$, $2\pi\psi_0$, otherwise stochastic effects are expected to come into play~\cite{Assadullahi:2016gkk, Vennin:2016wnk}. 

Finally, let us mention that in Appendix~\ref{sec:inhom}, it is analytically shown that the difference between the forward and the backward formulations $\frac{3}{5}(\fnl^\mathrm{forward}-\fnl^\mathrm{backward})$ is given by $\fnl^\mathrm{CR}=\frac{1-\nS(k_\uS)}{4}$, for generic $V(\phi)$ and $\phi_C(\psi)$ functions. This further supports the consistency of our two formulations.
\section{Conclusions}
\label{sec:conclusion}
Future galaxy surveys should bring the non-Gaussianity (NG) detection threshold down to $\fnl\sim 1$. In this context, the practical calculation of the bispectrum in multiple-field models of inflation is an important task. The $\delta N$ formalism provides such a calculational framework. In its standard implementation however, it does not allow one to go beyond the near-equilateral limit of the squeezed bispectrum, and it requires to compute its intrinsic component separately from the third order action. Furthermore, it has recently been pointed out that when specified to what would be measured by a local observer, an additional correction proportional to $1-\nS$ should be taken into account.

In this paper, we have introduced a new approach to the calculation of the squeezed bispectrum in the $\delta N$ formalism. In this framework, as well as in the recently developed soft limit expansion~\cite{Kenton:2015lxa, Byrnes:2015dub, Kenton:2016abp}, the intrinsic NG is automatically included as a slow-roll correction so that the calculation of $\fnl$ does not rely on anything more than solving background dynamics, and the difference between the small and the large wavelengths is accounted for which allows one to go beyond the near-equilateral limit. In the examples we discussed, it was shown that the near-equilateral limit is indeed not always a valid estimate for more generic configurations. Finally, our approach can be formulated in terms of comoving scales (``forward'' formulation) or in terms of physical scales as seen by a local observer (``backward'' formulation). In the first case, the standard results are recovered in the near-equilateral limit but in the second case, the $1-\nS$ correction mentioned above is included. In particular, the squeezed $\fnl$ parameter measured by a local observer vanishes in single-field inflation.

Although the explicit formulas given for $\fnl$ in the present work rely on the slow-roll approximation, our formalism can be applied to more generic setups as soon as the system has relaxed to a phase-space attractor. Let us also mention that at the level of the bispectrum, this work confirms that the local observer effect leads to a cancellation of Maldacena's consistency relation in single-field inflation. At the level of the trispectrum, the Suyama-Yamaguchi inequality~\cite{Suyama:2007bg} provides another consistency relation in the forward formulation. How it changes when expressed in the backward formulation, \ie once the local observer effect is taken into account, would be worth studying.  
Finally, in the case where the quantum diffusion of the inflationary fields substantially affects their dynamics, 
the $\delta N$ formalism has recently been extended to the stochastic-$\delta N$ formalism~\cite{Enqvist:2008kt, Fujita:2013cna, 
Fujita:2014tja, Vennin:2015hra, Assadullahi:2016gkk, Vennin:2016wnk, Kawasaki:2015ppx}, that takes these quantum backreaction effects into account. This work also paves the way for a calculation of the squeezed bispectrum in stochastic inflation.
\acknowledgments
It is a pleasure to thank Zachary Kenton, David J. Mulryne, 
David Wands and Shuichiro Yokoyama for interesting comments and enjoyable discussions.
This work is supported by STFC grants ST/K00090X/1 and ST/N000668/1,
and by the World Premier International Research Center Initiative (WPI), MEXT, Japan. 
Y.T. is supported by JSPS Research Fellowship for Young Scientists.
\begin{appendix}
\section{Practical recipe for a few classes of models}
In \Sec{sec:ComputationalProgramme}, we have explained how the squeezed $\fnl(\kL,\kS)$ parameter can be computed in generic multi-field models of inflation. In the limit where the two scales are equal, $\kL=\kS$, analytical formulas were provided. For the generic case $\kL<\kS$ however, one has to resort to numerical techniques in order to compute the correlators $\langle \zeta_\uL  \calP_\zeta(\kS)\rangle$. In this appendix, we detail the computational programme one has to follow for a few classes of models, concrete examples of which are provided in \Sec{sec:examples}. Generalisation to other models than those presented here can be done along the same lines. We focus on the backward formulation, while
the calculations in the forward formulation are briefly mentioned in footnotes.
\subsection{Additive separable potentials}
\label{sec:sepAddPot}
Let us first consider the case of a two-field additive separable potential
\bea
V\left(\phi,\psi\right)=U\left(\phi\right)+W\left(\psi\right).
\eea
The two fields $\phi$ and $\psi$ are assumed to be slowly rolling during inflation, and evolve according to
\bea
\frac{\dd\phi}{\dd N} = -\Mp^2\frac{U^\prime}{V}, \quad\quad\quad
\frac{\dd\psi}{\dd N} = -\Mp^2\frac{W^\prime}{V}.
\label{eq:additive:KG}
\eea
Additive separable potentials are convenient to work with since the following two formulas can be derived. 
First, there is an integral of motion~\cite{Sasaki:2008uc}
\begin{align}
\label{K:def}
K\left(\phi,\psi\right)=\int^\phi\dfrac{\dd\tilde{\phi}}{U^\prime(\tilde{\phi})}-\int^\psi\dfrac{\dd\tilde{\psi}}{W^\prime(\tilde{\psi})},
\end{align}
which allows us to label different slow-roll trajectories. Indeed, if one differentiates $K$ with respect to $N$ and make use of \Eq{eq:additive:KG}, one can readily show that $K$ is a constant. Second, if two points $M_1(\phi_1,\psi_1)$ and $M_2(\phi_2,\psi_2)$ are on the same slow-roll trajectory 
(that is to say, $K(\phi_1,\psi_1)=K(\phi_2,\psi_2)$), the number of \efolds realised between $M_1$ and $M_2$ is given by~\cite{Starobinsky:1986fxa}
\begin{align}
\label{eq:efolds}
N_{M_1 M_2} = -\dfrac{1}{\Mp^2}\int_{\phi_1}^{\phi_2}\dfrac{U}{U^\prime}\dd\phi
-\dfrac{1}{\Mp^2}\int_{\psi_1}^{\psi_2}\dfrac{W}{W^\prime}\dd\psi\, .
\end{align}
Let us now fix the three level lines $\rho=\rho_\uc$, $N_{\rightarrow\mathrm{c}}=N_\uS $ and $N_{\rightarrow\mathrm{c}}=N_\uL $ as in
\Fig{fig:schema}. Let $A_\uin(\phi,\psi)$ be a free point in field space, and $B$ and $C$ the points associated to $A_\uin$ according to \Fig{fig:schema}. Let us calculate the derivatives of the coordinates of $B$ and $C$ and of $N=N_{AC}$ with respect to $\phi$ and $\psi$. These will be useful to calculate $\fnl$.

The values of $\phi_{C}$ and $\psi_{C}$ are the same for all the points belonging to the same slow-roll trajectory, hence they depend only on $K$. One then has
\begin{align}
\label{eq:dphicdphi:dphicdK:additive}
	\begin{cases}
		\displaystyle
		\pdif{\phi_{C}}{\phi} = \pdif{K}{\phi}\dif{\phi_{C}}{K}, & 
		\displaystyle
		\pdif{\phi_{C}}{\psi} = \pdif{K}{\psi}\dif{\phi_{C}}{K}, \\[15pt]
		\displaystyle
		\pdif{\psi_{C}}{\phi} = \pdif{K}{\phi}\dif{\psi_{C}}{K}, & 
		\displaystyle
		\pdif{\psi_{C}}{\psi} = \pdif{K}{\psi}\dif{\psi_{C}}{K},
	\end{cases}
\end{align}
where, differentiating \Eq{K:def}, one has $\partial K/\partial\phi = 1/U^\prime(\phi)$ and $\partial K/\partial\psi = -1/W^\prime(\psi)$. 
Differentiating the condition $U(\phi_{C})+W(\psi_{C})=\rho_{C}$, one also has $U_{C}^\prime \dd\phi_{C}/\dd K + W_{C}^\prime \dd\psi_{C}/\dd K = 0$. 
On the other hand, differentiating $K(\phi_{C},\psi_{C})$ given in \Eq{K:def} with respect to $K$, one has $1=(\dd\phi_{C}/\dd K)/U^\prime_{C}-(\dd\psi_{C}/\dd K)/W^\prime_{C}$. 
These two relations give rise to
\begin{align}
	\dif{\phi_{C}}{K} = \frac{1}{U_{C}^\prime}\left(\frac{1}{{U^\prime_{C}}^2}+\frac{1}{{W^\prime_{C}}^2}\right)^{-1},
	\quad
	\dif{\psi_{C}}{K} = -\frac{1}{W_{C}^\prime}\left(\frac{1}{{U^\prime_{C}}^2}+\frac{1}{{W^\prime_{C}}^2}\right)^{-1}.
\end{align}
Combining these results, one obtains
\begin{align}\label{C derivative add}
	\begin{cases}
		\displaystyle
		\pdif{\phi_{C}}{\phi} = \frac{1}{U^\prime(\phi)}
		\frac{U^\prime_{C} {W_C^\prime}^2}{{U_C^\prime}^2+{W_C^\prime}^2}, & 
		\displaystyle
		\pdif{\phi_C}{\psi} = -\frac{1}{W^\prime(\psi)}
		\frac{U^\prime_C {W_C^\prime}^2}{{U_C^\prime}^2+{W_C^\prime}^2}, \\[15pt]
		\displaystyle
		\pdif{\psi_C}{\phi} = -\frac{1}{U^\prime(\phi)}
		\frac{{U^\prime_C}^2 {W_C^\prime}}{{U_C^\prime}^2+{W_C^\prime}^2}, & 
		\displaystyle
		\pdif{\psi_C}{\psi} = \frac{1}{W^\prime(\psi)}
		\frac{{U^\prime_C}^2 {W_C^\prime}}{{U_C^\prime}^2+{W_C^\prime}^2}.
	\end{cases}
\end{align}

The coordinates of $B$ are defined through the two conditions $K(\phi_B,\psi_{B})=K(\phi,\psi)$ and $N_{BC}=N_\uS $. 
By differentiating these two equations with respect to $\phi$ and $\psi$, one obtains a linear system of four equations, from which, 
making use of \Eqs{C derivative add}, the four following quantities can be extracted\footnote{In the forward formulation,
the coordinates of $B$ are defined through the two conditions $K(\phi_B,\psi_B)=K(\phi,\psi)$ and $N_{AB}=N_\uL-N_\uS$.
By differentiating these two equations with respect to $\phi$ and $\psi$, one obtains~\cite{Kenton:2015lxa}
\begin{align}
	\begin{cases}
		\displaystyle
		\pdif{\phi_B}{\phi} = \frac{U_B^\prime}{U^\prime(\phi)}\frac{U(\phi)+W_B}{V_B}, &
		\displaystyle
		\pdif{\phi_B}{\psi} = \frac{U_B^\prime}{W^\prime(\psi)}\frac{W(\psi)-W_B}{V_B}, \\[15pt]
		\displaystyle
		\pdif{\psi_B}{\phi} = \frac{W_B^\prime}{U^\prime(\phi)}\frac{U(\phi)-U_B}{V_B}, &
		\displaystyle
		\pdif{\psi_B}{\psi} = \frac{W_B^\prime}{W^\prime(\psi)}\frac{W(\psi)+U_B}{V_B}.
	\end{cases}
\end{align}
The other required derivatives can be obtained in the same way as with the backward formulation.
}
\begin{align}
\label{eq:dphiBdphi:additive}
	\begin{cases}
		\displaystyle
		\pdif{\phi_{B}}{\phi} = \frac{U_B^\prime}{U^\prime(\phi)V_B}
		\left(\frac{U_C {W_C^\prime}^2-W_C {U_C^\prime}^2}{{U_C^\prime}^2+{W_C^\prime}^2}+W_B\right), & 
		\displaystyle
		\pdif{\phi_{B}}{\psi} = \frac{U_B^\prime}{W^\prime(\psi)V_B}
		\left(\frac{W_C {U_C^\prime}^2-U_C {W_C^\prime}^2}{{U_C^\prime}^2+{W_C^\prime}^2}-W_B\right), \\[15pt]
		\displaystyle
		\pdif{\psi_{B}}{\phi} = \frac{W_B^\prime}{U^\prime(\phi)V_B}
		\left(\frac{U_C {W_C^\prime}^2-W_C {U_C^\prime}^2}{{U_C^\prime}^2+{W_C^\prime}^2}-U_B\right), &
		\displaystyle
		\pdif{\psi_{B}}{\psi} = \frac{W_B^\prime}{W^\prime(\psi)V_B}
		\left(\frac{W_C {U_C^\prime}^2-U_C {W_C^\prime}^2}{{U_C^\prime}^2+{W_C^\prime}^2}+U_B\right).
	\end{cases}
\end{align}

Then, to calculate the long-wavelength curvature perturbation $\zeta_\uL =\delta N_\uL $, the derivatives of the backward number of \efolds are needed.
Differentiating $N=N_{AC}$ given by \Eq{eq:efolds} (where $M_1=A$ and $M_2=C$) with respect to $\phi$ and $\psi$, one obtains
\begin{align}
	\pdif{N}{\phi} = \frac{1}{\Mp^2}\left[\frac{U(\phi)}{U^\prime(\phi)}-\frac{W_C}{W_C^\prime}\pdif{\psi_C}{\phi}
	-\frac{U_C}{U_C^\prime}\pdif{\phi_C}{\phi}\right], \quad
	\pdif{N}{\psi} = \frac{1}{\Mp^2}\left[\frac{W(\psi)}{W^\prime(\psi)}-\frac{W_C}{W_C^\prime}\pdif{\psi_C}{\psi}
	-\frac{U_C}{U_C^\prime}\frac{\partial\phi_C}{\partial\psi}\right],
\end{align}
which, combined with \Eq{C derivative add}, gives rise to
\begin{align}
\label{eq:dNdphi:additive}
	\pdif{N}{\phi} = \frac{1}{\Mp^2  U^\prime(\phi)} 
	\left[U(\phi)+\frac{W_C {U^\prime}^2_C - U_C {W_C^\prime}^2}{{U_C^\prime}^2 + {W_C^\prime}^2}\right], \quad
	\pdif{N}{\psi} = \frac{1}{\Mp^2  W^\prime(\psi)} 
	\left[W(\psi)+\frac{U_C {W^\prime_C}^2 - W_C {U_C^\prime}^2}{{U_C^\prime}^2 + {W_C^\prime}^2}\right].
\end{align}
In the following, the second derivatives of $N$ are also needed. They can be obtained by differentiating the previous expressions one more time and making use of \Eq{C derivative add} again. One obtains
\begin{align}
	\hspace{-20pt}
	\left\{
	\begin{aligned}
		\Mp^2\pdif{{}^2N}{\phi^2} =&\, 1 - \frac{U^{\prime\prime}(\phi)}{{U^\prime}^2(\phi)}
		\left[U(\phi)+\frac{W_C {U^\prime_C}^2 - U_C {W_C^\prime}^2}{{U_C^\prime}^2 + {W_C^\prime}^2}\right] 
		+\frac{1}{{U^\prime}^2(\phi)}\frac{{U_C^\prime}^2{W_C^\prime}^2}{\left({U_C^\prime}^2+{W_C^\prime}^2\right)^2}
		\\ 	& \times
		\left[2\left(U_C^{\prime\prime}W_C+U_C W_C^{\prime\prime}\right)- {U_C^\prime}^2-{W_C^\prime}^2 
		 +2\left(U_C^{\prime\prime}+W_C^{\prime\prime}\right)\frac{U_C {W^\prime_C}^2 - W_C {U_C^\prime}^2}{{U_C^\prime}^2 
		+ {W_C^\prime}^2}\right], \\
		\Mp^2\pdif{{}^2 N}{\psi^2} =&\, 1 - \frac{W^{\prime\prime}(\psi)}{{W^\prime}^2(\psi)}
		\left[W(\psi)+\frac{U_C {W^\prime_C}^2 - W_C {U_C^\prime}^2}{{U_C^\prime}^2 + {W_C^\prime}^2}\right] 
		+\frac{1}{{W^\prime}^2(\psi)}\frac{{U_C^\prime}^2{W_C^\prime}^2}{\left({U_C^\prime}^2+{W_C^\prime}^2\right)^2}
		\\ & \times
		\left[2\left(U_C^{\prime\prime}W_C+U_C W_C^{\prime\prime}\right)- {U_C^\prime}^2-{W_C^\prime}^2 
		+2\left(U_C^{\prime\prime}+W_C^{\prime\prime}\right)\frac{U_C {W^\prime_C}^2 - W_C {U_C^\prime}^2}{{U_C^\prime}^2 
		+ {W_C^\prime}^2}\right], \\
		\Mp^2\ppdif{N}{\phi}{\psi} =&\, \frac{2}{U^\prime(\phi)W^\prime(\psi)}
		\frac{{U_C^\prime}^2{W_C^\prime}^2}{\left({U_C^\prime}^2+{W_C^\prime}^2\right)^2} \\
		&\times\left[{U_C^\prime}^2+{W_C^\prime}^2-2\left(U_C^{\prime\prime}W_C+U_C W_C^{\prime\prime}\right)
		+2\left(U_C^{\prime\prime}+W_C^{\prime\prime}\right)
		\frac{W_C {U^\prime_C}^2 - U_C {W_C^\prime}^2}{{U_C^\prime}^2 + {W_C^\prime}^2}\right].
	\end{aligned}
	\right.
\end{align}

The power spectrum of the scalar curvature perturbations realised between $A(\phi,\psi)$ and the constant energy hypersurface 
$\rho=\rho_\uc$ is given by
\bea
\label{eq:Pzeta:additive}
\calP_\zeta\left(\phi,\psi\right)=\left[\left(\dfrac{\partial N}{\partial\phi}\right)^2+\left(\dfrac{\partial N}{\partial\psi}\right)^2\right]\left(\dfrac{H}{2\pi}\right)^2,
\eea
where $H^2=V/(3\Mp^2)$ and $\partial N/\partial\phi$ and and $\partial N/\partial\psi$ are given in \Eqs{eq:dNdphi:additive}. This gives rise to 
\begin{align}
\label{eq:dPzetadphi:additive}
	\begin{cases}
		\displaystyle
		6\Mp^2\pi^2\pdif{\calP_\zeta}{\phi} =\left(\pdif{N}{\phi}\pdif{{}^2 N}{\phi^2}+\pdif{N}{\psi}\ppdif{N}{\phi}{\psi}\right)V(\phi,\psi)
		+\frac{1}{2}\left[\left(\pdif{N}{\phi}\right)^2+\left(\pdif{N}{\psi}\right)^2\right]U^\prime(\phi), \\[15pt]
		\displaystyle
		6\Mp^2\pi^2\pdif{\calP_\zeta}{\psi} = \left(\pdif{N}{\psi}\pdif{{}^2 N}{\psi^2}+\pdif{N}{\phi}\ppdif{N}{\phi}{\psi}\right)V(\phi,\psi)
		+\frac{1}{2}\left[\left(\pdif{N}{\phi}\right)^2+\left(\pdif{N}{\psi}\right)^2\right]W^\prime(\psi).
	\end{cases}
\end{align}
All the quantities required to evaluate \Eq{fNL beyond equilateral} have now been specified, and $\fnl$ can be computed. In practice, starting from $A_\uin$, the following computational program should be used
\begin{enumerate}
\item \label{enum:additive:1} Compute the coordinates in field space of $A_*$ by numerically solving $K(A_\uin)=K(A_*)$ and $N_{A_*C}=N_\uL $, where $K$ is given by \Eq{K:def} and $N$ by \Eq{eq:efolds}
\item  \label{enum:additive:2} Compute the coordinates of $B$ by numerically solving $K(A_\uin)=K(B)$ and $N_{BC}=N_\uS $, where $K$ is given by \Eq{K:def} and $N$ by \Eq{eq:efolds}
\item  \label{enum:additive:3} Evaluate $\left.\calP_\zeta\right\vert_*$ making use of \Eq{eq:Pzeta:additive} and of the result of step~\ref{enum:additive:1}
\item   \label{enum:additive:4} Evaluate $\left.\calP_\zeta\right\vert_B$ making use of \Eq{eq:Pzeta:additive} and of the result of step~\ref{enum:additive:2}
\item   \label{enum:additive:5} Evaluate $\left.\partial\calP_\zeta/\partial\phi^i\right\vert_B$ making use of \Eq{eq:dPzetadphi:additive} and of the result of step~\ref{enum:additive:2}
\item \label{enum:additive:6} Evaluate $\left.\partial\phi^i_B/\partial\phi^j\right\vert_*$ making use of \Eq{eq:dphiBdphi:additive} and of the result of step~\ref{enum:additive:1}
\item Evaluate \Eq{fNL beyond equilateral} with the results of steps~\ref{enum:additive:3}-\ref{enum:additive:6}.
\end{enumerate}

Finally, let us notice that the case where $U=W$ is effectively equivalent to a single-field setup. Specifying the previous formulas in this case, 
it is easy to see that $\partial\phi_B/\partial\phi=-\partial\psi_B/\partial\phi=-\partial\psi_B/\partial\phi=\partial\psi_B/\partial\psi$, 
and that $\partial\calP_\zeta/\partial\phi=\partial\calP_\zeta/\partial\psi$. From here it follows that $\fnl=0$ in this case, as it should. 
\subsection{Multiplicative separable potentials}
\label{sec:sepMulPot}
Let us know derive the same computational program but in the case of a multiplicative separable potential
\begin{align}
V\left(\phi,\psi\right)=U\left(\phi\right)W\left(\psi\right).
\end{align}
The two fields are still assumed to be slowly rolling during inflation, according to
\bea
\frac{\partial\phi}{\partial N}=-\Mp^2\frac{U^\prime}{U}, \quad\quad\quad
\frac{\partial\psi}{\partial N}=-\Mp^2\frac{W^\prime}{W}.
\eea
In this case, an integral of motion is given by~\cite{Sasaki:2008uc}
\begin{align}
\label{K:mul:def}
K\left(\phi,\psi\right)=\int^\phi\dfrac{U(\tilde{\phi})}{U^\prime(\tilde{\phi})}\dd\tilde{\phi}-\int^\psi\dfrac{W(\tilde{\psi})}{W^\prime(\tilde{\psi})}\dd\tilde{\psi},
\end{align}
which labels different slow-roll trajectories. Then, if two points $M_1(\phi_1,\psi_1)$ and $M_2(\phi_2,\psi_2)$ are on the same slow-roll trajectory [that is to say, $K(\phi_1,\psi_1)=K(\phi_2,\psi_2)$], the number of \efolds realised between $M_1$ and $M_2$ is given by,
\begin{align}
\label{eq:efolds:Mult}
N_{M_1 M_2} = -\dfrac{1}{\Mp^2}\int_{\phi_1}^{\phi_2}\dfrac{U}{U^\prime}\dd\phi
=-\dfrac{1}{\Mp^2}\int_{\psi_1}^{\psi_2}\dfrac{W}{W^\prime}\dd\psi,
\end{align}
where either of the two expression can be used according to what is more convenient.
As in Appendix~\ref{sec:sepAddPot}, let us now fix the three level lines $\rho=\rho_\uc$, $N_{\rightarrow\mathrm{c}}=N_\uS $ and $N_{\rightarrow\mathrm{c}}=N_\uL $ as in
\Fig{fig:schema}. Let $A_\uin(\phi,\psi)$ be a free point in field space, and $B$ and $C$ the points associated to $A_\uin$ according to \Fig{fig:schema}.

The values of $\phi_C$ and $\psi_C$ are the same for all the points belonging to the same slow-roll trajectory, hence they depend only on $K$ and \Eqs{eq:dphicdphi:dphicdK:additive} also apply for multiplicative potentials. On the other hand, differentiating \Eq{K:mul:def}, one has $\partial K/\partial\phi = U(\phi)/U^\prime(\phi)$ and $\partial K/\partial\psi = -W(\psi)/W^\prime(\psi)$. Differentiating the condition $U(\phi_C)W(\psi_C)=\rho_\uc$, one also has $W_CU_C^\prime \dd\phi_C/\dd K + U_CW_C^\prime \dd\psi_C/\dd K = 0$, and differentiating $K(\phi_C,\psi_C)$ given in \Eq{K:mul:def} with respect to $K$, one obtains $1=(\dd\phi_C/\dd K) (U_C/U^\prime_C)-(\dd\psi_C/\dd K)(W_C/W^\prime_C)$. These two relations give rise to
\begin{align}
	\dif{\phi_C}{K} = \left(\frac{U_C}{U^\prime_C}
	+\frac{W_C^2U^\prime_C}{{W^\prime_C}^2U_C}\right)^{-1}, \quad
	\dif{\psi_C}{K} = -\left(\frac{W_C}{W^\prime_C}
	+\frac{U_C^2W^\prime_C}{{U^\prime_C}^2W_C}\right)^{-1},
\end{align}
and combining these results, one obtains
\begin{align}
	\label{C derivatives mul}
	\begin{cases}
		\displaystyle
		\pdif{\phi_C}{\phi} = \frac{U(\phi)}{U^\prime(\phi)}\left(\frac{U_C}{U^\prime_C}
		+\frac{W_C^2U^\prime_C}{{W^\prime_C}^2U_C}\right)^{-1}, &
		\displaystyle 
		\pdif{\phi_C}{\psi} = -\frac{W(\psi)}{W^\prime(\psi)}\left(\frac{U_C}{U^\prime_C}
		+\frac{W_C^2U^\prime_C}{{W^\prime_C}^2U_C}\right)^{-1}, \\[15pt]
		\displaystyle
		\pdif{\psi_C}{\phi} = -\frac{U(\phi)}{U^\prime(\phi)}\left(\dfrac{W_C}{W^\prime_C}
		+\frac{U_C^2W^\prime_C}{{U^\prime_C}^2W_C}\right)^{-1}, &
		\displaystyle
		\pdif{\psi_C}{\psi} = \frac{W(\psi)}{W^\prime(\psi)}\left(\dfrac{W_C}{W^\prime_C}
		+\frac{U_C^2W^\prime_C}{{U^\prime_C}^2W_C}\right)^{-1}.
	\end{cases}
\end{align}

The coordinates of $B$ are defined through the two conditions $K(\phi_B,\psi_B)=K(\phi,\psi)$ and $N_{BC}=N_\uS $. 
By differentiating these two equations with respect to $\phi$ and $\psi$, one obtains a linear system of four equations, from which, 
making use of the above formulas, the four following quantities can be extracted\footnote{In the forward formulation,
the coordinates of $B$ are defined through $K(\phi_B,\psi_B)=K(\phi,\psi)$ and $N_{AB}=N_\uL-N_\uS$, giving
\begin{align}
	\pdif{\phi_B}{\phi}=\frac{U(\phi)}{U^\prime(\phi)}\frac{U_B^\prime}{U_B}, \quad
	\pdif{\psi_B}{\psi}=\frac{W(\psi)}{W^\prime(\psi)}\frac{W_B^\prime}{W_B}, \quad 
	\pdif{\phi_B}{\psi}=\pdif{\psi_B}{\phi}=0.
\end{align}
The other derivatives can be obtained in the same way as with the backward formulation.
}
\begin{align}
\label{eq:dphiBdphi:multiplicative}
	\begin{cases}
		\displaystyle
		\pdif{\phi_B}{\phi} = \frac{U(\phi)}{U^\prime(\phi)}\frac{U^\prime_B}{U_B}
		\left(1+\frac{W_C^2 {U^\prime_C}^2}{U_C^2 {W^\prime_C}^2}\right)^{-1}, &
		\displaystyle
		\pdif{\phi_B}{\psi} = -\frac{W(\psi)}{W^\prime(\psi)}\frac{U^\prime_B}{U_B}
		\left(1+\frac{W_C^2 {U^\prime_C}^2}{U_C^2 {W^\prime_C}^2}\right)^{-1}, \\[15pt]
		\displaystyle
		\pdif{\psi_B}{\phi} = -\frac{U(\phi)}{U^\prime(\phi)}\frac{W^\prime_B}{W_B}
		\left(1+\frac{U_C^2 {W^\prime_C}^2}{W_C^2 {U^\prime_C}^2}\right)^{-1}, &
		\displaystyle
		\pdif{\psi_B}{\psi} = \frac{W(\phi)}{W^\prime(\psi)}\frac{W^\prime_B}{W_B}
		\left(1+\frac{U_C^2 {W^\prime_C}^2}{W_C^2 {U^\prime_C}^2}\right)^{-1}.
	\end{cases}
\end{align}

Differentiating $N=N_{AC}$ given by \Eq{eq:efolds:Mult} (where $M_1=A$ and $M_2=C$) with respect to $\phi$ and $\psi$, one obtains
\begin{align}
	\begin{cases}
		\displaystyle
		\pdif{N}{\phi} = \frac{1}{\Mp^2}\left[\frac{U(\phi)}{U^\prime(\phi)}-\frac{U_C}{U_C^\prime}\pdif{\phi_C}{\phi}\right]
		= -\frac{1}{\Mp^2}\pdif{\psi_C}{\phi}\frac{W_C}{W^\prime_C}, \\[15pt]
		\displaystyle
		\pdif{N}{\psi} = \frac{1}{\Mp^2}\left[\frac{W(\psi)}{W^\prime(\psi)}-\frac{W_C}{W_C^\prime}\pdif{\psi_C}{\psi}\right]
		= -\frac{1}{\Mp^2}\pdif{\phi_C}{\psi}\frac{U_C}{U^\prime_C}.
	\end{cases}
\end{align}
Making use of \Eq{C derivatives mul}, this gives rise to
\begin{align}
	\dfrac{\partial N}{\partial \phi} =\dfrac{1}{\Mp^2}\dfrac{U(\phi)}{ U^\prime(\phi)} 
	\left(1+\dfrac{U_C^2{W^\prime_C}^2}{{W_C}^2{U_C^\prime}^2}\right)^{-1}, \quad
	\dfrac{\partial N}{\partial \psi} =\dfrac{1}{\Mp^2}\dfrac{W(\psi)}{ W^\prime(\psi)} 
	\left(1+\dfrac{W_C^2{U^\prime_C}^2}{{U_C}^2{W_C^\prime}^2}\right)^{-1}.
\end{align}
The second derivatives of $N$ are also needed. They can be obtained by differentiating the previous expressions one more time and making use of \Eq{C derivatives mul} again. One obtains, after a few manipulations,
\begin{align}
	\left\{
	\begin{aligned}
		\displaystyle
		\Mp^2\dfrac{\partial^2 N}{\partial\phi^2} = &
		\displaystyle 
		\left[1-\dfrac{U(\phi)U^{\prime\prime}(\phi)}{{U^\prime}^2(\phi)}\right]
		\left(1+\dfrac{U_C^2{W_C^\prime}^2}{W_C^2{U_C^\prime}^2}\right)^{-1} \\
		& \displaystyle
		-2\dfrac{U^2(\phi)}{{U^\prime}^2(\phi)}\dfrac{U_C^2W_C^2{U_C^\prime}^4{W_C^\prime}^4}
		{\left(U_C^2{W_C^\prime}^2+W_C^2{U_C^\prime}^2\right)^{3}}\left(2-\dfrac{U_C U_C^{\prime\prime}}
		{{U_C^\prime}^2}-\dfrac{W_C W_C^{\prime\prime}}{{W_C^\prime}^2}\right), \\
		\displaystyle
		\Mp^2\dfrac{\partial^2 N}{\partial\psi^2} = & 
		\displaystyle
		\left[1-\dfrac{W(\psi)W^{\prime\prime}(\psi)}{{W^\prime}^2(\psi)}\right]
		\left(1+\dfrac{W_C^2{U_C^\prime}^2}{U_C^2{W_C^\prime}^2}\right)^{-1} \\
		&\displaystyle
		-2\dfrac{W^2(\psi)}{{W^\prime}^2(\psi)}\dfrac{U_C^2W_C^2{U_C^\prime}^4{W_C^\prime}^4}
		{\left(U_C^2{W_C^\prime}^2+W_C^2{U_C^\prime}^2\right)^{3}}\left(2-\dfrac{U_C U_C^{\prime\prime}}
		{{U_C^\prime}^2}-\dfrac{W_C W_C^{\prime\prime}}{{W_C^\prime}^2}\right), \\
		\displaystyle
		\Mp^2\dfrac{\partial^2 N}{\partial\phi\partial\psi} = &
		\displaystyle 
		2\dfrac{U(\phi)W(\psi)}{U^\prime(\phi)W^\prime(\psi)}
		\dfrac{U_C^2W_C^2{U_C^\prime}^4{W_C^\prime}^4}{\left(U_C^2{W_C^\prime}^2+W_C^2{U_C^\prime}^2\right)^{3}}
		\left(2-\dfrac{U_C U_C^{\prime\prime}}{{U_C^\prime}^2}-\dfrac{W_C W_C^{\prime\prime}}{{W_C^\prime}^2}\right).
	\end{aligned}
	\right.
\end{align}

The power spectrum of the scalar curvature perturbations realised between $A(\phi,\psi)$ and the constant energy hypersurface $\rho=\rho_\uc$ 
is given by
\begin{align}
\label{eq:Pzeta:multiplicative}
\calP_\zeta\left(\phi,\psi\right)=\left[\left(\dfrac{\partial N}{\partial\phi}\right)^2+\left(\dfrac{\partial N}{\partial\psi}\right)^2\right]\left(\dfrac{H}{2\pi}\right)^2,
\end{align}
where $H^2=UW/(3\Mp^2)$. This gives rise to 
\begin{align}
\label{eq:dPzetadphi:multiplicative}
	\begin{cases}
		\displaystyle
		6\Mp^2\pi^2\dfrac{\partial\calP_\zeta}{\partial\phi} =\left(\dfrac{\partial N}{\partial\phi}
		\dfrac{\partial^2 N}{\partial\phi^2}+\dfrac{\partial N}{\partial\psi}\dfrac{\partial^2 N}{\partial\phi\partial\psi}\right)U(\phi)W(\psi)
		+\dfrac{1}{2}\left[\left(\dfrac{\partial N}{\partial\phi}\right)^2+\left(\dfrac{\partial N}{\partial\psi}\right)^2\right]U^\prime(\phi)W(\psi), \\[15pt]
		\displaystyle
		6\Mp^2\pi^2\dfrac{\partial\calP_\zeta}{\partial\psi} =\left(\dfrac{\partial N}{\partial\psi}
		\dfrac{\partial^2 N}{\partial\psi^2}+\dfrac{\partial N}{\partial\phi}\dfrac{\partial^2 N}{\partial\phi\partial\psi}\right)U(\phi)W(\psi)
		+\dfrac{1}{2}\left[\left(\dfrac{\partial N}{\partial\phi}\right)^2+\left(\dfrac{\partial N}{\partial\psi}\right)^2\right]U(\phi)W^\prime(\psi).
	\end{cases}
\end{align}
From these expressions, a computational programme similar to the one given in Appendix~\ref{sec:sepAddPot} can be obtained:
\begin{enumerate}
\item \label{enum:multiplicative:1} Compute the coordinates in field space of $A_*$ by numerically solving $K(A_\uin)=K(A_*)$ and $N_{A*C}=N_\uL $, where $K$ is given by \Eq{K:mul:def} and $N$ by \Eq{eq:efolds:Mult}
\item  \label{enum:multiplicative:2} Compute the coordinates of $B$ by numerically solving $K(A_\uin)=K(B)$ and $N_{BC}=N_\uS $, where $K$ is given by \Eq{K:mul:def} and $N$ by \Eq{eq:efolds:Mult}
\item  \label{enum:multiplicative:3} Evaluate $\left.\calP_\zeta\right\vert_*$ making use of \Eq{eq:Pzeta:multiplicative} and of the result of step~\ref{enum:multiplicative:1}
\item   \label{enum:multiplicative:4} Evaluate $\left.\calP_\zeta\right\vert_B$ making use of \Eq{eq:Pzeta:multiplicative} and of the result of step~\ref{enum:multiplicative:2}
\item   \label{enum:multiplicative:5} Evaluate $\left.\partial\calP_\zeta/\partial\phi^i\right\vert_B$ making use of \Eq{eq:dPzetadphi:multiplicative} and of the result of step~\ref{enum:multiplicative:2}
\item \label{enum:multiplicative:6} Evaluate $\left.\partial\phi^i_B/\partial\phi^j\right\vert_*$ making use of \Eq{eq:dphiBdphi:multiplicative} and of the result of step~\ref{enum:multiplicative:1}
\item Evaluate \Eq{fNL beyond equilateral} with the results of steps~\ref{enum:multiplicative:3}-\ref{enum:multiplicative:6}.
\end{enumerate}
\subsection{Inhomogeneous end of inflation}
\label{sec:inhom}
In this section we consider the case where inflation is driven by a single scalar field $\phi$ through the potential $V(\phi)$, but ends at a value $\phi_C(\psi)$ that depends on the value of an additional scalar field $\psi$. In this case, the number of \efolds realised between $M_1$ and $M_2$ is given by
\bea
N_{M_1M_2} = -\dfrac{1}{\Mp^2}\int_{\phi_1}^{\phi_2}\dfrac{V}{V^\prime}\dd\phi,
\label{eq:NM1M2:inhom}
\eea
provided $\psi_1=\psi_2$. According to the terminology used in Appendices~\ref{sec:sepAddPot} and~\ref{sec:sepMulPot}, $K=\psi$ is an integral of motion.

One simply has $\phi_C(\phi,\psi)=\phi_C(\psi)$ and $\psi_C(\phi,\psi)=\psi$ since $\psi$ does not evolve, hence
\bea
	\dfrac{\partial\phi_C}{\partial\phi} = 0, \quad \dfrac{\partial\phi_C}{\partial\psi} = \phi_C^\prime(\psi), \quad
	\dfrac{\partial\psi_C}{\partial\phi} = 0, \quad \dfrac{\partial\psi_C}{\partial\psi} = 1.
\eea
The coordinates of $B$ are then defined through the two conditions $\psi_B=\psi$ and  $N_{BC}=N_\uS $. 
By differentiating these expressions with respect to $\phi$ and $\psi$, this gives rise to\footnote{In the forward formulation,
the coordinates of $B$ are defined through $\psi_B=\psi$ and $N_{AB}=N_\uL-N_\uS$, giving
\begin{align}
	\pdif{\phi_B}{\phi}=\frac{V(\phi)}{V^\prime(\phi)}\frac{V_B^\prime}{V_B}, \quad
	\pdif{\psi_B}{\psi}=1, \quad \pdif{\phi_B}{\psi}=\pdif{\psi_B}{\phi}=0.
\end{align}
The other derivatives can be obtained in the same way as with the backward formulation.
}
\bea
\label{eq:dphiBdphi:inhomogeneous}
	\dfrac{\partial\phi_B}{\partial \phi} = 0, \quad 
	\dfrac{\partial\phi_B}{\partial\psi} = \dfrac{V_C}{V^\prime_C}\dfrac{V^\prime_B}{V_B}\phi_C^\prime(\psi), \quad
	\dfrac{\partial\psi_B}{\partial \phi} = 0, \quad
	\dfrac{\partial\psi_B}{\partial\psi}  = 1.
\eea 

Differentiating $N=N_{AC}$ given by \Eq{eq:NM1M2:inhom} (where $M_1=A$ and $M_2=C$) with respect to $\phi$ and $\psi$, one obtains
\bea
\dfrac{\partial N}{\partial\phi} = \dfrac{1}{\Mp^2}\dfrac{V(\phi)}{V^\prime(\phi)}, \quad\quad\quad
\dfrac{\partial N}{\partial\psi} = -\dfrac{1}{\Mp^2}\dfrac{V_C}{V^\prime_C}\phi_C^\prime(\psi),
\eea
and the second derivatives can be obtained by differentiating these expressions, giving rise to
\bea
	\begin{cases}
		\displaystyle
		\dfrac{\partial^2 N}{\partial\phi ^2} = \dfrac{1}{\Mp^2}\left[1-\dfrac{V(\phi)V^{\prime\prime}(\phi)}{{V^\prime}^2(\phi)}\right], \\[15pt]
		\displaystyle
		\dfrac{\partial^2 N}{\partial\psi ^2} = -\dfrac{1}{\Mp^2}\left[{\phi_C^\prime}^2\left(\psi\right) 
		\left(1-\dfrac{V_C V_C^{\prime\prime}}{{V_C^\prime}^2}\right) +\dfrac{V_C}{V^\prime_C}\phi_C^{\prime\prime}
		\left(\psi\right)\right], \\[15pt]
		\displaystyle
		\dfrac{\partial^2 N}{\partial\phi \partial\psi} = 0.
	\end{cases}
\eea

The power spectrum of the scalar curvature perturbations realised between $A(\phi,\psi)$ and the surface $\phi=\phi_C(\psi)$ is given by 
\bea
\label{eq:Pzeta:inhomogeneous}
	\calP_\zeta\left(\phi,\psi\right)=\left[\left(\dfrac{\partial N}{\partial\phi}\right)^2
	+\left(\dfrac{\partial N}{\partial\psi}\right)^2\right]\left(\dfrac{H}{2\pi}\right)^2
	= \dfrac{V(\phi)}{12\pi^2\Mp^6}\left[\dfrac{V^2(\phi)}{{V^\prime}^2(\phi)}+\left(\dfrac{V_C}{V^\prime_C}\phi^\prime_C\right)^2\right],
\eea
and its derivatives can then be calculated according to
\bea
\label{eq:dPzetadphi:inhomogeneous}
	\begin{cases}
		\displaystyle
		12\pi^2\Mp^6 \dfrac{\partial\calP_\zeta}{\partial\phi} = 
		3\dfrac{V^2(\phi)}{V^\prime(\phi)}-2\dfrac{V^3(\phi)V^{\prime\prime}(\phi)}{{V^\prime}^3(\phi)}
		+V^\prime(\phi)\left(\dfrac{V_C}{V^\prime_C}\phi^\prime_C\right)^2, \\[15pt]
		\displaystyle
		12\pi^2\Mp^6 \dfrac{\partial\calP_\zeta}{\partial\psi} = 
		2V(\phi) \dfrac{V_C}{V^\prime_C}\phi_C^\prime
		\left[\left(1-\dfrac{V_C V_C^{\prime\prime}}{{V_C^\prime}^2}\right){\phi_C^\prime}^2+\dfrac{V_C}{V_C^\prime}
		\phi_C^{\prime\prime}\right].
	\end{cases}
\eea
Starting from $A_\uin$, the computational program is therefore given by
\begin{enumerate}
\item \label{enum:inhomogeneous:1} Compute $\phi_*$ by numerically solving $N_{A*C}=N_\uL $, where $N$ is given by \Eq{eq:NM1M2:inhom}
\item  \label{enum:inhomogeneous:2} Compute $\phi_B$ by numerically solving $N_{BC}=N_\uS $, where $N$ is given by \Eq{eq:NM1M2:inhom}
\item  \label{enum:inhomogeneous:3} Evaluate $\left.\calP_\zeta\right\vert_*$ making use of \Eq{eq:Pzeta:inhomogeneous} and of the result of step~\ref{enum:inhomogeneous:1}
\item   \label{enum:inhomogeneous:4} Evaluate $\left.\calP_\zeta\right\vert_B$ making use of \Eq{eq:Pzeta:inhomogeneous} and of the result of step~\ref{enum:inhomogeneous:2}
\item   \label{enum:inhomogeneous:5} Evaluate $\left.\partial\calP_\zeta/\partial\phi^i\right\vert_B$ making use of \Eq{eq:dPzetadphi:inhomogeneous} and of the result of step~\ref{enum:inhomogeneous:2}
\item \label{enum:inhomogeneous:6} Evaluate $\left.\partial\phi^i_B/\partial\phi^j\right\vert_*$ making use of \Eq{eq:dphiBdphi:inhomogeneous} and of the result of step~\ref{enum:inhomogeneous:1}
\item Evaluate \Eq{fNL beyond equilateral} with the results of steps~\ref{enum:inhomogeneous:3}-\ref{enum:inhomogeneous:6}.
\end{enumerate}

Finally let us note that, in this model, one can analytically show that the difference between the forward and the backward formulations 
$\frac{3}{5}(\fnl^\mathrm{forward}-\fnl^\mathrm{backward})$ is equal to the CR component $\fnl^\mathrm{CR}=\frac{1-\nS(k_\uS)}{4}$, 
irrespectively of the concrete expressions of $V(\phi)$ and $\phi_C(\psi)$, since
\begin{align}
	\frac{3}{5}(\fnl^\mathrm{forward}-\fnl^\mathrm{backward})=\frac{1-\nS(k_\uS)}{4}=\frac{3V_\uB^2{V_\uB^\prime}^2{V_\uC^\prime}^2
	+{V_\uB^\prime}^4V_\uC^2{\phi_\uC^\prime}^2-2V_\uB^3V_\uB^{\prime\prime}{V_\uC^\prime}^2}{4(V_\uB^4{V_\uC^\prime}^2
	+V_\uB^2{V_\uB^\prime}^2V_\uC^2{\phi_\uC^\prime}^2)}\Mp^2.
\end{align}
\end{appendix}
\bibliographystyle{JHEP}
\bibliography{classical}
\end{document}